\pdfoutput=1
\documentclass[12pt, draftclsnofoot, onecolumn]{IEEEtran} 
\usepackage{multirow}
\usepackage{diagbox}
\usepackage{amssymb}
\usepackage{amsmath}
\usepackage{graphicx}
\usepackage{cite}
\usepackage{citesort}
\usepackage{subfigure}
\usepackage{graphicx,epstopdf}
\usepackage{epsfig}	
\usepackage{cite,graphicx,amsmath,amssymb}
\usepackage{comment}
\usepackage{amssymb}
\usepackage{amsmath}
\usepackage{cite}
\usepackage{url}
\usepackage{xcolor}
\usepackage{cite,graphicx,amsmath,amssymb}
\usepackage{subfigure}
\usepackage{citesort}
\usepackage{fancyhdr}
\usepackage{mdwmath}
\usepackage{mdwtab}
\usepackage{caption}
\usepackage{amsthm}
\usepackage{lipsum}
\usepackage{adjustbox}

\newtheorem{theorem}{Theorem}

\newtheorem{lemma}{Lemma}

\newtheorem{corollary}{Corollary}

\newtheorem{remark}{Remark}  
\newtheorem{proposition}{Proposition}

\makeatletter
\def\ScaleIfNeeded{%
\ifdim\Gin@nat@width>\linewidth \linewidth \else \Gin@nat@width
\fi } \makeatother

\UseRawInputEncoding

\begin{document}
\title{\Huge{Active Simultaneously Transmitting and Reflecting Surface Assisted NOMA Networks}}

\author{Xinwei Yue,~\IEEEmembership{Senior Member,~IEEE}, Jin Xie, Chongjun Ouyang, Yuanwei Liu,~\IEEEmembership{Senior Member,~IEEE}, Xia Shen and  Zhiguo~Ding,~\IEEEmembership{Fellow, IEEE}

\thanks{X. Yue and J. Xie are with the Key Laboratory of Information and Communication Systems, Ministry of Information Industry and also with the Key Laboratory of Modern Measurement $\&$ Control Technology, Ministry of Education, Beijing Information Science and Technology University, Beijing 100101, China (email: \{xinwei.yue and jin.xie\}@bistu.edu.cn).}
\thanks{Chongjun Ouyang is with the School of Information and Communication Engineering, Beijing University of Posts and Telecommunications, Beijing, 100876, China (e-mail: dragonaim@bupt.edu.cn).}
\thanks{Y. Liu is with the School of Electronic Engineering and Computer Science, Queen Mary University of London, London E1 4NS, U.K. (email: yuanwei.liu@qmul.ac.uk).}
\thanks{X. Shen is with the China Academy of Information and Communications Technology (CAICT), Beijing 100191, China (email: shenxia@caict.ac.cn).}
\thanks{Z. Ding is with the Department of Electrical Engineering And Computer Science, Khalifa University, Abu Dhabi, UAE, and also with Department of Electrical Engineering, Princeton University, Princeton, USA (e-mail: zhiguo.ding@gmail.com).}
}
\maketitle

\begin{abstract}
The novel active simultaneously transmitting and reflecting surface (ASTARS) has recently received a lot of attention due to its capability to conquer the multiplicative fading loss and achieve full-space smart radio environments. This paper introduces the ASTARS to assist non-orthogonal multiple access (NOMA) communications, where the stochastic geometry theory is used to model the spatial positions of pairing users. We design the independent reflection/transmission phase-shift controllers of ASTARS to align the phases of cascaded channels at pairing users. We derive new closed-form and asymptotic expressions of the outage probability and ergodic data rate for ASTARS-NOMA networks in the presence of perfect/imperfect successive interference cancellation (pSIC). The diversity orders and multiplexing gains for ASTARS-NOMA are derived to provide more insights. Furthermore, the system throughputs of ASTARS-NOMA are investigated in both delay-tolerant and delay-limited transmission modes. The numerical results are presented and show that: 1) ASTARS-NOMA with pSIC outperforms ASTARS assisted-orthogonal multiple access (ASTARS-OMA) in terms of outage probability and ergodic data rate; 2) The outage probability of ASTARS-NOMA can be further reduced within a certain range by increasing the power amplification factors; 3) The system throughputs of ASTARS-NOMA are superior to that of ASTARS-OMA in both delay-limited and delay-tolerant transmission modes.
\end{abstract}
\section{Introduction}
As the number of device in wireless networks explosively grows, the sixth-generation (6G) wireless networks are facing unprecedented challenges for providing high-speed, low-latency data services for massive users \cite{To6G2,To6G1}.
From the standpoint of expanding capacity, signal strength, and coverage range, the reconfigurable intelligent surface (RIS) boasts remarkable capabilities. It has been considered as one of promising technologies of 6G networks \cite{RIS1}. Essentially, passive RIS (PRIS) is a planar surface comprising abundant inexpensive passive reflecting components, which is able to modify the phase and amplitude of incident signals to achieve smart radio environments \cite{RIS3}. In addition, PRIS has been shown to be able to improve the performance of physical layer security \cite{Yang2020RISSecrecy}, user localization\cite{RISpositioning} and unmanned aerial vehicle communications \cite{RISUAV}.

Despite of the aforementioned advantages of PRIS, it only provides the half-space smart transmissions \cite{RIS2,LIRS}. The innovative simultaneously transmitting and reflecting surfaces (STARS), which can achieve full spatial coverage, was proposed to get around this restriction \cite{STARDi}. Specifically, a passive STARS (PSTARS) integrates many passive simultaneously transmitting and reflecting elements that can transmit and reflect the incident signals \cite{STARliang}. Based on hardware architecture and physical principles, the authors in \cite{2021STAR} further studied mode switching, energy splitting and time switching protocols for PSTARS networks.
For these three protocols, the authors of \cite{TWCSTARMu} investigated the minimization of power consumption in PSTARS networks.
The impact of separate codebooks on PSTARS was examined in terms of detection capabilities and system performance without complete channel state information (CSI)\cite{STARsong}.
To enhance channel conditions, the authors in \cite{STARair} incorporated PSTARS into the over-the-air computation system, enabling excellent learning accuracy and privacy perservation  over large coverage areas.
In \cite{STARXu2}, the authors employed a diversity-preserving phase-shift strategy to attain complete diversity order of PSTARS networks by taking into account coupled phase-shift models.
As a further development, the authors in \cite{STARSecur} evaluated the secrecy capacity of PSTARS networks with coupled phase-shift scheme.

Widespread interests have also been drawn to non-orthogonal multiple access (NOMA) which is a potential multiple access method for the next generation of wireless communication networks \cite{NGMAIOT,NGMAliu}. NOMA has the ability to boost system throughput, capacity and energy efficiency in comparison to orthogonal multiple access (OMA), delivering an improved communication service for large numbers of users \cite{Yuan2021NOMA6G}. The notion of cooperative NOMA was introduced in \cite{Ding2014Cooperative}, where one cell-centred user is utilized as a relay to enhance the quality of service for an edge user. Inspired by this work, the ergodic data rate and outage probability of full/half-duplex cooperative NOMA networks were studied in \cite{Yue8026173}. The authors integrated PRIS into NOMA networks \cite{Ding2020Shifting}, where the effect of stochastic discrete and coherent phase-shifting designs was researched for PRIS-NOMA networks. With the focus on green communications, the authors of \cite{FangRISNOMA} revealed the tradeoff between maximizing the sum rate and minimizing the power budget in PRIS-assisted NOMA networks. Considering complexity expansion and error propagation issues, the authors in \cite{Yue2020IRSNOMA} studied the ergodic data rate and outage performances of PRIS-NOMA with perfect/imperfect successive interference cancellation (pSIC/ipSIC) schemes.
Recently, a new concept of near-field NOMA communication was introduced in \cite{nearNOMA}, which benefits from the beamforming characteristics of near-field to enable NOMA in both angular and distance domains. Moreover, the authors of \cite{nearNOMADing} utilized pre-configured spatial beams to serve both near-field and far-field users, confirming that NOMA can effectively support the coexistence of near-field and far-field communications.

As mentioned above, the integration of NOMA with other technologies is flourishing, and the PSTARS-assisted NOMA (PSTARS-NOMA) networks naturally becomes a promising direction. The superiority of NOMA related on the differentiated channel conditions among users \cite{DobreNOMA}, and thus the establishment of channel condition differences was essential for NOMA. With the help of PSTARS, the users can be deployed to different half-spaces with vastly disparate channel conditions, thereby augmenting the performance of NOMA \cite{MASTAR1}. From the perspective of performance analysis, the outage performance of PSTARS-NOMA was evaluated by utilizing the central limit theorem and curve fitting model \cite{ChaoIOS}. On the basis of these models, the authors of \cite{STARS-NOMA} analyzed the ergodic data rate, outage behaviors and system throughput of PSTARS-NOMA with pSIC/ipSIC schemes. The coverage characteristics of PSTARS networks were surveyed in \cite{STARCoverage}, where the coverage of PSTARS-NOMA can be significantly extended compared to PSTARS-OMA.
In the presence of Nakagami-$m$ cascade channels, the secrecy outage probability of PSTARS-NOMA was researched in \cite {SOPSTAR} by considering the residual hardware impairments.
The authors of \cite{STARRA} researched a matching theory based channels allocation scheme to achieve the maximum sum rate of PSATRS-NOMA systems. In \cite{WLSTAR1}, over-the-air federated learning and PSTARS-NOMA were integrated into an unified framework, which achieves both high spectral efficiency and learning performance.

While PRIS/PSTARS bring the enhanced performance of wireless networks, they also cause multiplicative fading loss. Specifically, the small-scale fading of transmitter-PSTARS/PRIS link and PSTARS/PRIS-receiver link were multiplied, which is usually worse than direct-link fading \cite{ARIS6G}. To eliminate this effect, an active RIS (ARIS) with integrated reflection-type amplifiers has been proposed \cite{LongARIS}, which magnifies the signals' power, and then reflect to the desired users. The simulation results demonstrated that the service area coverage and spectral efficiency of ARIS were superior to those of PRIS \cite{ARISBasar}. Condition on the same power consumption, the authors of \cite{PanARIS} revealed that ARIS outperforms PRIS in terms of the achievable data rate if the number of elements is small. In energy-constrained internet-of-things systems, ARIS-NOMA was proven to achieve higher system throughput than ARIS-OMA \cite{ChenARIS}. A subarray-based ARIS structure was designed to improve energy efficiency \cite{ZhuARIS}, where each subarray can be independently controlled. Recently, a novel hardware model for active STARS (ASTARS) was proposed \cite{JiaqiARIS}, which has the ability to offset the multiplicative fading loss and achieve full-space coverage. In \cite{MaARIS}, the authors confirmed that ASTRAS-aided communication systems outperform ARIS in terms of the sum-rate improvement and power consumption reduction. Moreover, the maximum secrecy rate of ASTARS assisted wireless networks was achieved by jointly optimizing the configuration of elements and beamforming of access points \cite{GouARIS}.

\subsection{Motivation and Contributions}
As a new topic, only a few works have been researched for ASTARS networks, where the hardware model design \cite{JiaqiARIS}, sum rate maximisation \cite{MaARIS} and system security \cite{GouARIS} have been the focus of the previous works. Since ASTARS is able to provide different channel differences and amplify the desired signals for non-orthogonal users, the physical layer performance analysis of ASTARS assisted NOMA networks is necessary to gain valuable insights. To the best of our knowledge, the integration of ASTARS with NOMA networks have not been researched yet, and the critical questions require further exploration. In particularly, considering the issues of complexity scaling and error propagation, it is important to analyse the effect of ipSIC on ASTARS-NOMA networks. The impact of the ASTARS elements' configuration affects on the performance of ASTARS-NOMA networks is still unknown. Inspired by these motivations, we introduce an ASTARS to assist NOMA communications by invoking stochastic geometry, where the pairing users, i.e., $U_r$ and $U_t$ are randomly distributed within contralateral area of ASTARS. More particularly, we evaluate the outage probability, system throughput, and ergodic data rate for $U_r$ with pSIC/ipSIC and $U_t$. In summary, the following are the primary contributions of this paper:
\begin{enumerate}
  \item We propose ASTARS-NOMA networks with randomly deployed pairing users, where $U_r$ and $U_t$ are located at the opposite sides of ASTARS for NOMA transmission. We design the independent reflection/transmission phase-shift controllers of ASTARS to align the phases of the cascaded channels at $U_r$ and $U_t$, respectively. We derive the closed-form expressions of the outage probability for $U_r$ with pSIC/ipSIC and $U_t$ by invoking the stochastic geometry. We also investigate the system throughput of ASTARS-NOMA in the delay-limited transmission mode.
  \item We derive the asymptotic expressions of the outage probability for $U_r$ with pSIC/ipSIC and $U_t$ by utilising Laplace transforms and convolution theorem. The diversity orders of $U_r$ with pSIC/ipSIC and $U_t$ in the high SNR region are calculated, respectively. We confirm that the diversity orders of $U_r$ with pSIC and $U_t$ are proportional to the quantities of ASTARS elements. The outage probability of $U_r$ with ipSIC converges to an error floor due to the residual interference, and the corresponding diversity order is equal to $zero$.
  \item We derive the closed-form expressions of ergodic data rate for $U_r$ with pSIC/ipSIC and $U_t$. We further derive asymptotic expressions of ergodic data rate for $U_r$ with ipSIC and $U_t$ within high SNR region. Based on Jensen's inequality, we provide an upper bound on $U_r$'s ergodic data rate with pSIC. On the basis of approximated analyses, we survey the multiplexing gains for $U_r$ and $U_t$. Moreover, the system throughputs of ASTARS-NOMA are evaluated in the delay-tolerant transmission mode.
  \item We compare the performance of ASTARS-NOMA with ASTARS-OMA and PSTARS-NOMA in terms of the outage probability, system throughput, and ergodic data rate. We reveal that both the outage probability and ergodic data rate of ASTARS-NOMA with pSIC performs better than ASTARS-OMA. On the condition of equipping with less ASTARS elements, ASTARS-NOMA is capable of furnishing the enhanced performance relative to PSTARS-NOMA. We further demonstrate that the outage behaviors of ASTARS-NOMA can be further improved within a certain range by increasing the power amplification factors.
\end{enumerate}
\subsection{Organization and Notations}
The rest of this article is divided into the following sections. Section \ref{SystemModel} presents the system model of ASTARS-NOMA in terms of hardware architecture, network deployment, and channel statistics. The outage probability expressions of ASTARS-NOMA are derived in Section \ref{Outage Probability}, in which the diversity orders for $U_r$ and $U_t$ are provided. Section \ref{Ergodic Rate} evaluates the ergodic data rate of $U_r$ and $U_t$. The simulation results and the corresponding analyses are presented in Section \ref{Numerical Results}. Then the conclusions of this paper are given in Section \ref{Conclusion1}, and Appendix contains a collection of mathematical proofs.

The main symbols used in this article are as follows: The probability density function (PDF) of a random variable $X$ is denoted as ${f_X}\left(  \cdot  \right)$, and its cumulative distribution function (CDF) is denoted as ${F_X}\left(  \cdot  \right)$. $\mathbb{E}\{\cdot\}$ denotes the expectation and $\mathbb{D}\{\cdot\}$ denotes variance operations; ${\left(  \cdot  \right)^H}$ stands for conjugate-transpose operation.
\section{System-Model}\label{SystemModel}
\begin{figure}[htbp]
\centering
\subfigure[Network devices and users deployment.]{\label{System-Model}
\begin{minipage}[t]{0.5\linewidth} 
\centering
\includegraphics[width=0.9\textwidth,height=0.6\textwidth]{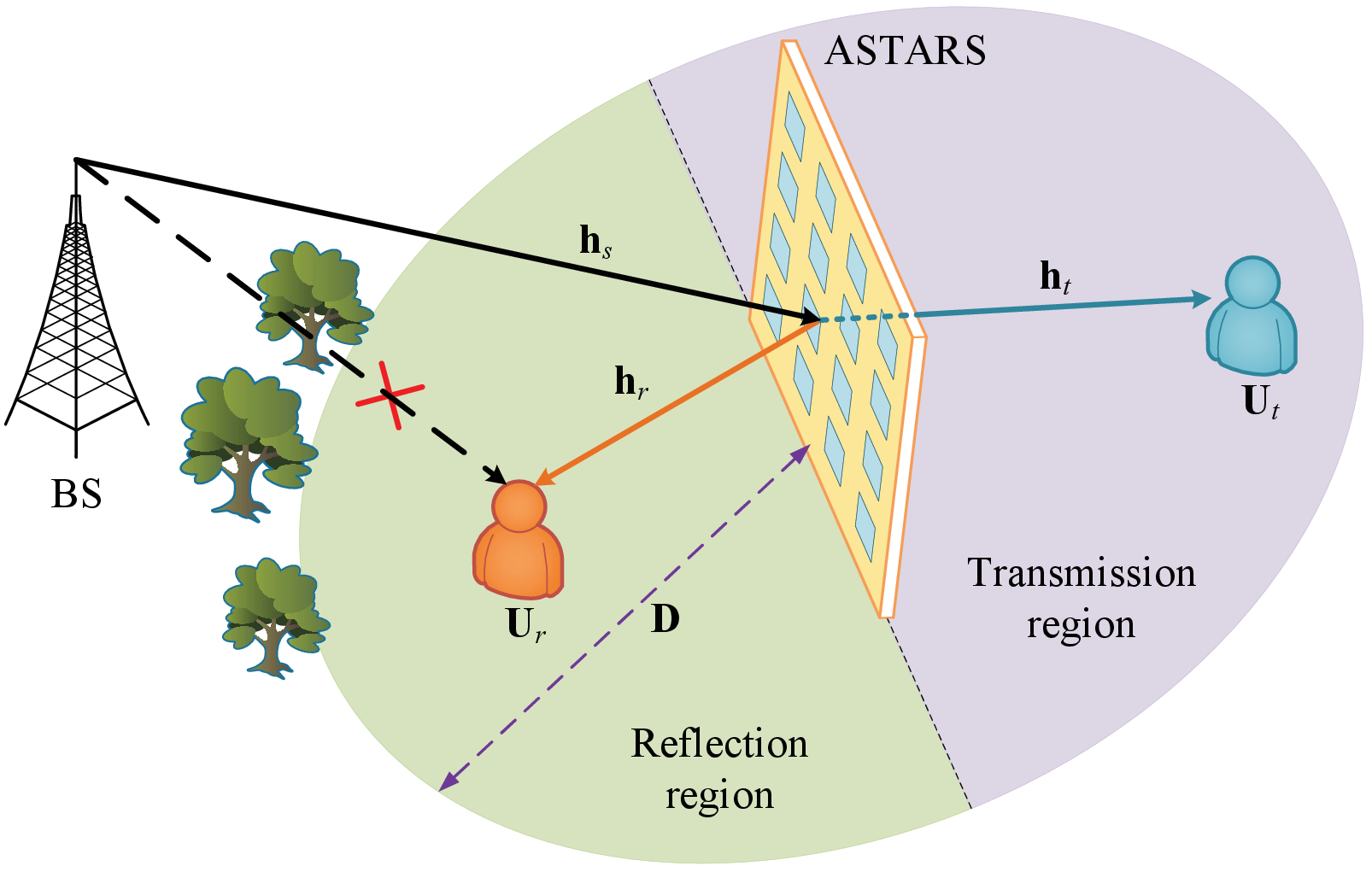}
\end{minipage}%
}%
\subfigure[Hardware structure of an ASTARS element.]{\label{element}
\begin{minipage}[t]{0.5\linewidth} 
\centering
\includegraphics[width=0.9\textwidth,height=0.6\textwidth]{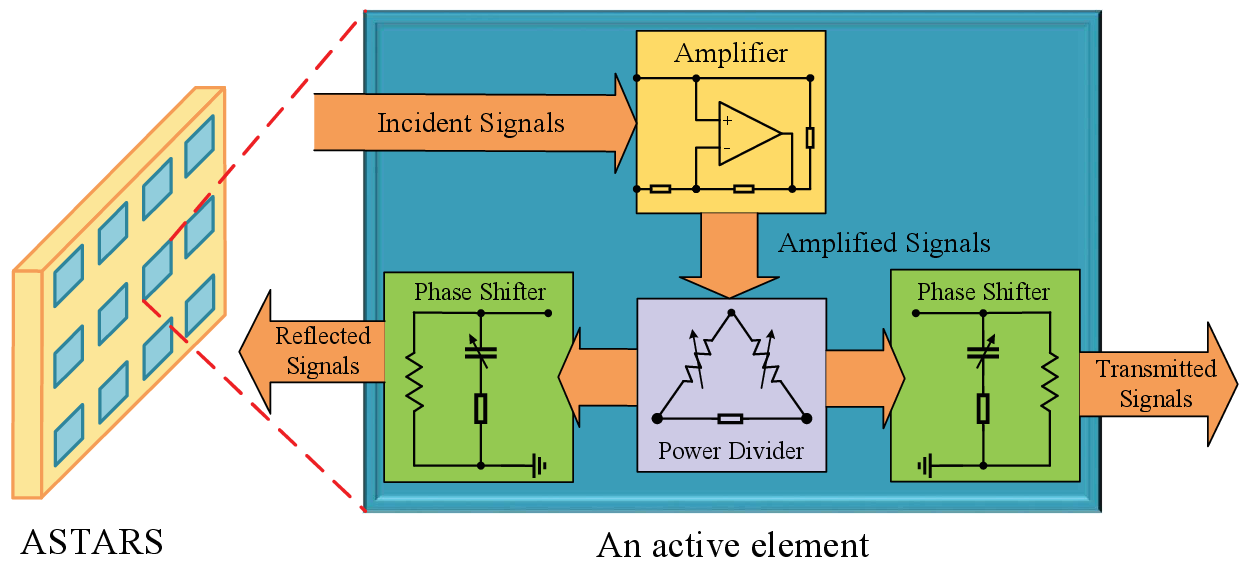}
\end{minipage}%
}%
\centering
\caption{System model of ASTARS-NOMA networks.}
\end{figure}

We consider an ASTARS-aided downlink NOMA communication scenario as illustrated in Fig. \ref{System-Model}, in which the incident signals from the base station (BS) are amplified and reflected or refracted to the users. The paring users and BS are both equipped with a single antenna, while the ASTARS is made up of $L$ ASTARS elements. Assume that the users are stochastically distributed in a circular region $\mathbb{O}_D$ with radius $D$, and ASTARS is fixed in the center of $\mathbb{O}_D$. More precisely, this circular region is separated by ASTARS into two parts denoted by the reflection region and transmission region, respectively. Two users are randomly selected from the reflection and transmission regions, and are denoted by $U_r$ and $U_t$, respectively. Due to the influence of obstacle blockage, assuming that the direct link from BS to $U_r$ and $U_t$ are not available or even in a state of complete outage. From the perspective of hardware design shown in Fig. 2, the element of  ASTRAS integrates active amplifiers for enlarging the incident signals to overcome the attenuation effect of multiplicative fading. Such integrated amplifiers can be implemented with many existing active devices, such as integrated chips \cite{ASTARS-T1}, the asymmetric current mirror \cite{ASTARS-T2} or the current-inverting converter \cite{ASTARS-T4}, which can significantly improve energy and hardware efficiency. Power splitting can be achieved by the harmonic components of power system \cite{ASTARS-T5}. The complex channel coefficients from the BS to ASTARS, and then from ASTARS to $U_\varphi$ are denoted by ${{\bf{h}}_{s}} \in \mathbb{C}{^{L \times 1}}$ and ${{{\bf{h}}_{\varphi} }} \in \mathbb{C}{^{L \times 1}}$ with $\varphi  \in \left\{ {r,t} \right\}$, respectively. For practical considerations, the ASTARS-NOMA networks' wireless communication links undergo Rician fading. ${{\bf{h}}_{r}^H{{\bf {\Theta}} _r}{{\bf{h}}_{s}}}$ and ${{\bf{h}}_{t}^H{{\bf {\Theta}} _t}{{\bf{h}}_{s}}}$ separately stand for the cascade complex channel coefficients from the BS to ASTARS, and then to ${U_r}$ and ${U_t}$, where ${{\bf {\Theta}} _r} = \sqrt {\lambda {\beta _r}} {\rm{diag}}\left( {{e^{j\theta _1^r}},...,{e^{j\theta _l^r}},...,{e^{j\theta _L^r}}} \right)= \sqrt {\lambda {\beta _r}} {{\bf{\Phi}} _r}$ and ${{\bf {\Theta}} _t} = \sqrt {\lambda {\beta _t}} {\rm{diag}}\left( {{e^{j\theta _1^t}},...,{e^{j\theta _l^t}},...,{e^{j\theta _L^t}}} \right)= \sqrt {\lambda {\beta _t}} {{\bf{\Phi}} _t}$ denote the reflection and transmission phase-shifting amplification matrixes of ASTARS, respectively. To facilitate analysis, assume that all ASTARS elements have the same amplification factor $\lambda $ and $\lambda  > 1$. ${{\beta _r}}$ and ${{\beta _t}}$ are denoted by the reflection and transmission amplitude coefficients, where ${\beta _r} + {\beta _t} \le 1$. ${\theta _l^r}$, ${\theta _l^t} \in \left[ {0,2\pi } \right)$ represent the transmission and reflection response's phase-shift of the $l$-th element, respectively. Since the transmission and reflection phase shifts are controlled by two different phase-shifters, ${\theta _l^r}$ and ${\theta _l^t}$ can be tuned independently. The perfect CSI is required for the users to carry out coherent demodulation.

\subsection{Signal Model of Pairing Users}
In ASTARS-NOMA networks, channels differences between users can be created by adjusting the energy coefficients of reflection and transmission \cite{ChaoIOS}. In this paper, we assign more energy coefficient to the users in reflection area, resulting in $U_r$ being the strong user and $U_t$ being the weak user. Based on the principle of NOMA, $U_r$ carries out the SIC process. To be specific, $U_r$ and $U_t$ receive the amplified superimposed signals reflected/transmitted from ASTARS and thermal noise generated by active components. At this moment, the received signal expressions at $U_r$ and $U_t$ can be separately written as
\begin{align}\label{The received signals of user n}
{y_r} = {\bf{h}}_r^H{{\bf{\Theta }}_r}{{\bf{h}}_s}\sqrt {P_s^{act}} {X_\Sigma } + {\bf{h}}_r^H{{\bf{\Theta }}_r}{{\bf{n}}_s} + {{\tilde n}_r},
\end{align}
and
\begin{align}\label{The received signals of user m}
{y_t} = {\bf{h}}_t^H{{\bf{\Theta }}_t}{{\bf{h}}_s}\sqrt {P_s^{act}} {X_\Sigma } + {\bf{h}}_t^H{{\bf{\Theta }}_t}{{\bf{n}}_s} + {{\tilde n}_t},
\end{align}
where ${X_\Sigma } = \sqrt {{a_r}} {x_r} + \sqrt {{a_t}} {x_t}$, ${{P_s^{act}}}$ denote BS's transmit power, ${x_r}$ and ${x_t}$ denote the signals of $U_r$ and $U_t$, respectively. $a_r$ and $a_t$ stands for the power allocation factor of $U_r$ and $U_t$, respectively. For the sake of fairness, $a_r$ and $a_t$ satisfy the relation ${a_r} \le {a_t}$ and ${a_r}+{a_t} = 1$. ${{\bf{n}}_s} = {\left[ {n_s^1, \cdots ,n_s^l, \cdots ,n_s^L} \right]^H}$ is denoted by the thermal noise matrix generated by ASTARS elements and ${n_s^l} \sim {\cal C}{\cal N}\left( {0,\sigma _s^2} \right)$. ${{\tilde n}_\varphi} \sim{\cal C}{\cal N} \left( {0,\sigma _0^2} \right)$ stands for white Gaussian noise with average power ${\sigma _0^2}$.
Let ${{\bf{h}}_s} = \sqrt {{\eta _0}d_s^{ - \alpha }} {\left[ {h_s^1, \cdots ,h_s^l, \cdots ,h_s^L} \right]^H}$, ${{\bf{h}}_\varphi } = \sqrt {{\eta _0}d_\varphi ^{ - \alpha }} {\left[ {h_\varphi ^1, \cdots ,h_\varphi ^l, \cdots ,h_\varphi ^L} \right]^H}$ denote the channel coefficients from the BS to ASTARS, and then from ASTARS to $U_\varphi$, where $h_s^l = {\sqrt {\frac{\kappa }{{\kappa  + 1}}}  + \sqrt {\frac{1}{{\kappa  + 1}}} \tilde h_s^l}$, $\tilde h_{\varphi }^l \sim {\cal C}{\cal N}\left( {0,1} \right)$, $h_\varphi ^l = {\sqrt {\frac{\kappa }{{\kappa  + 1}}}  + \sqrt {\frac{1}{{\kappa  + 1}}} \tilde h_\varphi ^l}$, $\tilde h_{s}^l \sim {\cal C}{\cal N}\left( {0,1} \right)$, $\kappa $ denotes the Rician factor and {$\alpha $ is path loss exponent, ${{\eta _0}}$ expresses the path loss, $d_{s}$ stands for the distances from BS to ASTARS, and $d_{\varphi}$ stand for the distance from ASTARS to $U_\varphi$. $U_r$ has better channel conditions and carried out the SIC to firstly detect the signal $x_t$ of $U_t$. Hence, the signal-plus-interference-to-noise ratio (SINR) for ${U_r}$ to decode $x_t$ can be expressed as
\begin{align}\label{SINR nm}
{\gamma _{r \to t}} = \frac{{{a_t}\lambda {\beta _r}P_s^{act}{{\left| {{\bf{h}}_r^H{{\bf{\Phi}} _r}{{\bf{h}}_s}} \right|}^2}}}{{{a_r}\lambda {\beta _r}P_s^{act}{{\left| {{\bf{h}}_r^H{{\bf{\Phi}} _r}{{\bf{h}}_s}} \right|}^2} + {\lambda {\beta _r}{{\left| {{\bf{h}}_r^H{{\bf{\Phi }}_r}{{\bf{n}}_s}} \right|}^2}} + \sigma _0^2}}.
\end{align}

After decoding and deleting $x_t$, the SNR for $U_r$ to decode its own information can be expressed as
\begin{align}\label{SINR n}
{\gamma _r} = \frac{{{a_r}\lambda {\beta _r}P_s^{act}{{\left| {{\bf{h}}_r^H{{\bf{\Phi }}_r}{{\bf{h}}_s}} \right|}^2}}}{{\lambda {\beta _r}{{\left| {{\bf{h}}_r^H{{\bf{\Phi }}_r}{{\bf{n}}_s}} \right|}^2} + \varepsilon {{\left| {{h_{re}}} \right|}^2}P_s^{act} + \sigma _0^2}},
\end{align}
where ${{{ {{h_{re}}} }}}  \sim {\cal C}{\cal N}\left( {0,{\sigma _{re}^2}} \right)$ stands for the residual interference caused by ipSIC. In particular, $\varepsilon {\rm{ = 0}}$ stands for pSIC and $\varepsilon {\rm{ = 1}}$ denotes ipSIC, respectively.

$U_t$ has weak channel conditions and thus regard $x_t$ and thermal noise as interference. The SINR for detecting $x_t$ can be given
by
\begin{align}\label{SINR m}
{\gamma _t} = \frac{{{a_t}\lambda {\beta _t}P_s^{act}{{\left| {{\bf{h}}_t^H{{\bf{\Phi}} _t}{{\bf{h}}_s}} \right|}^2}}}{{{a_r}\lambda {\beta _t}P_s^{act}{{\left| {{\bf{h}}_t^H{{\bf{\Phi}} _t}{{\bf{h}}_s}} \right|}^2} + \lambda {\beta _t}{{\left| {{\bf{h}}_t^H{{\bf{\Phi }}_t}{{\bf{n}}_s}} \right|}^2} + \sigma _0^2}}.
\end{align}
\subsection{Statistics Property of Channels}
The statistical characteristics of cascade Rician channels employed in ASTARS-NOMA networks are first provided in this subsection, and the spatial impacts on $U_r$ and $U_t$ are then evaluated.

\subsubsection{Cascade rician distribution} As ASTARS can independently control reflection and transmission phase-shifts, we configure ${\theta _l^r}$ and ${\theta _l^t}$ to align the cascaded channels' phase at $U_r$ and $U_t$, respectively. Thus the cascade Rician channels gain  ${{{\left| {{\bf{h}}_{\varphi}^H{{\bf{\Phi}} _\varphi }{{\bf{h}}_s }} \right|}^2}}$ can be rewritten as $\eta _0^2{\left( {{d_s}{d_\varphi }} \right)^{ - \alpha }}{\left| {\sum\nolimits_{l = 1}^L {\left| {h_s^lh_\varphi ^l} \right|} } \right|^2}$. Let ${{X_\varphi ^l} = \left| {h_s^lh_\varphi ^l} \right|}$ and $X_\varphi ={\left| {\sum\nolimits_{l = 1}^L {\left| {h_s^lh_\varphi ^l} \right|} } \right|^2}$. The PDF of  ${X_\varphi ^l}$ can be given by \cite{2006Probability}
\begin{align}\label{The CDF of cascade Rician channels}
{f_{X_\varphi ^l}}\left( x \right) = 4\sum\limits_{u = 0}^\infty  {\sum\limits_{v = 0}^\infty  {\frac{{{{\left( {\kappa  + 1} \right)}^{u + v + 2}}{x^{u + v + 1}}}}{{{{\left( {u!} \right)}^2}{{\left( v \right)}^2}{e^{2\kappa }}{\kappa ^{ - u - v}}}}} } {K_{u - v}}\left[ {2x\left( {\kappa  + 1} \right)} \right],
\end{align}
where ${K_x}\left(  \cdot  \right)$ indicates the modified Bessel function of the second kind with order $x$. We can separately express the mean and variance of $X_\varphi ^l$ as
\begin{align}\label{the mean of X_k}
\mathbb{E}\left( {{X_\varphi ^l}} \right) = \frac{{\pi }}{{4\left( {\kappa  + 1} \right)}}{\left[ {{L_{\frac{1}{2}}}\left( { - \kappa } \right)} \right]^2},
\end{align}
and
\begin{align}\label{the variance of X_k}
\mathbb{D}\left( {{X_\varphi ^l}} \right) =  {1 - \frac{{{\pi ^2}}}{{16{{\left( {\kappa  + 1} \right)}^2}}}{{\left[ {{L_{\frac{1}{2}}}\left( { - \kappa } \right)} \right]}^4}},
\end{align}
where ${L_{\frac{1}{2}}}\left( x  \right) = {e^{\frac{x}{2}}}\left[ {\left( {1 - x } \right){I_0}\left( { - \frac{x }{2}} \right) - x {I_1}\left( { - \frac{x }{2}} \right)} \right]$ is the Laguerre polynomial.

By applying Laguerre polynomial series \cite[Eq. (2.76)]{Primak2004}, the PDF and CDF for $X_\varphi$ are separately approximated as
\begin{align}\label{XPDF}
{f_{{X_\varphi }}}\left( x \right) = \frac{{{x^{\frac{{{p_\varphi }}}{2} - 1}}}}{{2q_\varphi ^{{p_\varphi }}\Gamma \left( {{p_\varphi }} \right)}}{e^{ - \frac{{\sqrt x }}{{{q_\varphi }}}}},
\end{align}
and
\begin{align}\label{XCDF}
{F_{{X_\varphi }}}\left( x \right) = \gamma \left( {{p_\varphi },\frac{{\sqrt x }}{{{q_\varphi }}}} \right)\Gamma {\left( {{p_\varphi }} \right)^{ - 1}},
\end{align} where $\gamma \left( {a,x} \right) = \int_0^x {{t^{a - 1}}{e^{ - t}}dt} $ is the lower incomplete Gamma function \cite[Eq. (8.350.1)]{2000gradshteyn} and $\Gamma \left(  \cdot  \right)$ is the gamma function \cite[Eq. (8.310.1)]{2000gradshteyn}, ${p_\varphi } = \frac{{K{\mathbb{E}^2}\left( {X_\varphi ^l} \right)}}{{\mathbb{D}\left( {X_\varphi ^l} \right)}}$,
${q_\varphi } = \frac{{\mathbb{D}\left( {X_\varphi ^l} \right)}}{{\mathbb{E}\left( {X_\varphi ^l} \right)}}$.

\subsubsection{Thermal noise intensity}\label{Thermal noise} Since ${\theta _l^\varphi}$ can only be phase-aligned with the ${{{\left| {{\bf{h}}_{\varphi}^H{{\bf{\Phi}} _\varphi }{{\bf{h}}_s }} \right|}^2}}$, the phase in each term of ${\left| {{\bf{h}}_{\varphi}^H{{\bf{\Phi }}_\varphi}{{\bf{n}}_s}} \right|^2} = {\eta _0}d_\varphi^{ - \alpha }\sigma _s^2{\left| {\sum\nolimits_{l = 1}^L {h_\varphi^l} } \right|^2}$ is considered to be randomly distributed. Let $H = {{\sum\nolimits_{l = 1}^L {h_\varphi^l} }}$, since $H$ is obtained by adding up $L$ independent identically distributed ${h_\varphi^l}$, it can be calculated as $H = L\sqrt {\frac{\kappa }{{\kappa  + 1}}}  + \sqrt {\frac{L}{{\kappa  + 1}}} \tilde H$, where $\tilde H$ follows a complex Gaussian distribution and $\tilde H \sim {\cal C}{\cal N}\left( {0,1} \right)$. We use the mean to characterize the channel power of $H$, which is expressed as $\mathbb{E}\left( {{{\left| H \right|}^2}} \right) = L\left( {\frac{{L\kappa  + 1}}{{\kappa  + 1}}} \right)$.

\subsubsection{User's location characteristics} For the path-loss experienced by the users, the PDFs of ${d_\varphi }$ can be obtained by using the fact that the locations
of paring users are stochastically distributed within ASTARS's serving area $\mathbb{O}_D$. In this case, the PDFs of $d_{r}$ and $d_{t}$ are written as \cite{ChaoIOS}
\begin{align}\label{dnPDF}
{f_{{d_r}}}\left( x \right) = \frac{\partial }{{\partial x}}\int_0^x {\int_0^\pi  {\frac{{2r}}{{\pi {D^2}}}} } drd\theta  = \frac{{2x}}{{{D^2}}},
\end{align}
and
\begin{align}\label{dmPDF}
{f_{{d_t}}}\left( x \right) = \frac{\partial }{{\partial x}}\int_0^x {\int_\pi ^{2\pi } {\frac{{2r}}{{\pi {D^2}}}} } drd\theta  = \frac{{2x}}{{{D^2}}},
\end{align}
respectively.
\section{Outage Probability}\label{Outage Probability}
In this section, the outage behaviors of ASTARS-NOMA are evaluated by invoking the stochastic geometry. To be more specific, the closed-form outage probability expressions of $U_r$ with pSIC/ipSIC and $U_t$ are derived for ASTARS-NOMA. To acquire further insight, the asymptotic expressions of outage probability and diversity orders for $U_r$ and $U_t$ are obtained.
\subsection{The $U_r$'s Outage Probability}
With the help of ASTARS, $U_r$, i.e., the user with strong channel conditions, needs to decode the information of $U_t$, and then decode its own signals. As a consequence,  the $U_r$ outage occurrences may be described as follows: 1) An outage event occurs when the signal $x_t$ of $U_t$ cannot be successfully decoded by $U_r$; and 2) The signal $x_t$ is successfully decoded, while the signal $x_r$ of $U_r$ fails to detect. Based on these explanations, the outage probability at $U_r$ in ASTARS-NOMA networks is shown as
\begin{align}\label{OP event n}
{P_{out,r}} = {\rm{Pr}}\left( {{\gamma _{r \to t}} > {{\hat \gamma }_t},{\gamma _r} < {{\hat \gamma }_r}} \right) + {\rm{Pr}}\left( {{\gamma _{r \to t}} < {{\hat \gamma }_t}} \right),
\end{align}
where ${{\hat \gamma }_r} = {2^{{{\hat R}_r}}} - 1$ and ${{\hat \gamma }_t} = {2^{{{\hat R}_t}}} - 1$ separately represent the target SNR for decoding $x_r$ and $x_t$. The corresponding target rates of $U_r$ and $U_t$ are defined as ${{{\hat R}_r}}$ and ${{{\hat R}_t}}$, respectively. The outage probability expression of $U_r$ with ipSIC for ASTARS-NOMA is illustrated in the following theorem.
\begin{theorem}\label{Theorem1:the OP of user n with ipSIC under Rician fading channel}
Condition on ${a_t} > {{\hat \gamma }_t}{a_r}$, the closed-form expression of $U_r$'s outage probability with ipSIC for ASTARS-NOMA is written as

\begin{align}\label{ipSIC}
P_{out,r}^{ipSIC}{\rm{ = }}\sum\limits_{k = 1}^K {\sum\limits_{u = 1}^U {\frac{{\pi {A_k}\left( {{x_u}{\rm{ + }}1} \right)}}{{2U\Gamma \left( {{p_r}} \right)}}\sqrt {1 - x_u^2} } }\gamma \left\{ {{p_r},\frac{1}{{{q_r}}}\sqrt {\frac{{{{\hat \gamma }_r}d_s^\alpha }}{{{a_r}P_s^{act}}}\left[ {\zeta \frac{{\sigma _s^2}}{{{\eta _0}}} + \frac{{\chi _u^\alpha }}{{\eta _0^2}}\left( {\frac{{\varepsilon {y_k}P_s^{act}}}{{{\beta _r}\lambda \sigma _{re}^{ - 2}}} + \frac{{\sigma _0^2}}{{{\beta _r}\lambda }}} \right)} \right]} } \right\},
\end{align}
where $\varepsilon  = 1$, $\zeta  = L\left( {\frac{{L\kappa  + 1}}{{\kappa  + 1}}} \right)$, ${x_u} = \cos \left( {\frac{{2u - 1}}{{2U}}\pi } \right)$, ${\chi _u} = \frac{{\left( {{x_u}{\rm{ + }}1} \right)D}}{2}$, ${{{ y}_k}}$ is the $k$-th zero point of Laguerre polynomial ${{ L}_{K}}\left( {{{ y}_k}} \right)$ and the $k$-th weight is expressed as ${A_k} = \frac{{{{\left( {K!} \right)}^2}{y_k}}}{{{{\left[ {{L_{K + 1}}\left( {{y_k}} \right)} \right]}^2}}}$. In addition, a trade-off between complexity and accuracy is also guaranteed by the parameters $K$ and $U$.
\begin{proof}
See Appendix~A.
\end{proof}
\end{theorem}
\begin{remark}\label{remark11}
If ${a_t} < {{\hat \gamma }_t}{a_r}$, and by substituting \eqref{SINR nm} and \eqref{SINR n} into \eqref{OP event n}, the expression of $U_r$'s outage probability with ipSIC can be written as
\begin{align}\label{inequality}
P_{out,r}^{ipSIC} = {\rm{Pr}}\left[ {{{\left| {{\bf{h}}_r^H{{\bf{\Phi }}_r}{{\bf{h}}_s}} \right|}^2} \ge \frac{\partial }{{P_s^{act}}}\left( {{{\left| {{\bf{h}}_r^H{{\bf{\Phi }}_r}{{\bf{n}}_s}} \right|}^2} + \frac{{\sigma _0^2}}{{{\beta _r}\lambda }}} \right)} \right],
\end{align}
where $\partial  = \frac{{{{\hat \gamma }_t}}}{{\left( {{a_t} - {{\hat \gamma }_t}{a_r}} \right)}}$.
Due to condition ${a_t} < {{\hat \gamma }_t}{a_r}$, the right side of the inequality \eqref{inequality} is less than zero, making the inequality always hold. At this time, the outage probability of ${U_r}$ with ipSIC will always equal to $one$.
\end{remark}

\begin{corollary}\label{Corollary:the OP of  user n with pSIC under Rician fading channel}
For case $\varepsilon =0$, the closed-form expression of $U_r$'s outage probability with pSIC for ASTARS-NOMA is written as
\begin{align}\label{pSIC}
P_{out,r}^{pSIC}=\sum\limits_{u = 1}^U {\gamma \left[ {{p_r},\frac{1}{{{q_r}}}\sqrt {\frac{{{{\hat \gamma }_r}d_s^\alpha }}{{{a_r}P_s^{act}}}\left( {\frac{{\chi _u^\alpha \sigma _0^2}}{{\eta _0^2{\beta _r}\lambda }} + \zeta \frac{{\sigma _s^2}}{{{\eta _0}}}} \right)} } \right]} \frac{{\pi \left( {{x_u}{\rm{ + }}1} \right)}}{{2U\Gamma \left( {{p_r}} \right)}}\sqrt {1 - x_u^2} .
\end{align}
\end{corollary}
\subsection{The $U_t$'s Outage Probability}
The following can be applied to indicate the outage occurrence at $U_t$: the SINR of decoded signal $x_t$ is lower the target SINR. The corresponding outage probability is shown as
\begin{align}\label{OP event m}
{P_{out,t}} = {\rm{Pr}}\left( {{\gamma _t} < {{\hat \gamma }_t}} \right).
\end{align}
\begin{theorem}\label{Theorem2:the OP of user m under Rician fading channel}
Condition on ${a_t} > {{\hat \gamma }_t}{a_r}$, the closed-form expression of $U_t$'s outage probability for ASTARS-NOMA is written as
\begin{align}\label{U_m}
{P_{out,t}} = \sum\limits_{u = 1}^U {\gamma \left[ {{p_t},\sqrt {\frac{{\partial d_s^\alpha q_t^{ - 2}}}{{P_s^{act}}}\left( {\frac{{y_u^\alpha \sigma _0^2}}{{\eta _0^2{\beta _t}\lambda }} + \zeta \frac{{\sigma _s^2}}{{{\eta _0}}}} \right)} } \right]} \frac{{\pi \left( {{x_u}{\rm{ + }}1} \right)}}{{2U\Gamma \left( {{p_t}} \right)}}\sqrt {1 - x_u^2} ,
\end{align}
where $\partial  = \frac{{{{\hat \gamma }_t}}}{{\left( {{a_t} - {{\hat \gamma }_t}{a_r}} \right)}}$. Similar to \textbf{Remark \ref{remark11}}, if ${a_t} < {{\hat \gamma }_t}{a_r}$, the outage probability of ${U_t}$ will always equal to $one$.

\begin{proof}
By substituting \eqref{SINR m} into \eqref{OP event m}, the outage probability expression of $U_t$ is further expressed as

\begin{align}
{P_{out,t}} ={\rm{Pr}}\left[ {{{\left| {{\bf{h}}_t^H{{\bf{\Phi }}_t}{{\bf{h}}_s}} \right|}^2} < \partial \left( {\frac{{{{\left| {{\bf{h}}_t^H{{\bf{\Phi }}_t}{{\bf{n}}_s}} \right|}^2}}}{{P_s^{act}}} + \frac{{\beta _t^{ - 1}\sigma _0^2}}{{P_s^{act}\lambda }}} \right)} \right].
\end{align}

By configuring the reflection phase-shift to align the phases of cascaded channels, the above expression can be rewritten as

\begin{align}
{P_{out,t}} = {\rm{Pr}}\left[ {{{\left| {\sum\limits_{l = 1}^L {h_s^lh_t^l} } \right|}^2} \le \frac{{\partial d_s^\alpha }}{{P_s^{act}}}\left( {\frac{{\zeta \sigma _s^2}}{{{\eta _0}}} + \frac{{d_t^\alpha \sigma _0^2}}{{\eta _0^2{\beta _t}\lambda }}} \right)} \right].
\end{align}

The following procedures resemble those in Appendix A. The proof is completed.
\end{proof}
\end{theorem}

\begin{proposition}\label{The system outage of STAR_RIS_NOMA}
When an outage occurs for at least one user in the system, it is considered as a system outage event. Hence, the ASTARS-NOMA's system outage probability with pSIC/ipSIC is written as
\begin{align}\label{system outage}
P_{NOMA,\tau }^{ASTARS} = 1 - \left( {1 - {P_{out,r}^{\tau }}} \right)\left( {1 - {P_{out,t}}} \right),
\end{align}
where  $\tau  \in \left\{ {ipSIC,pSIC} \right\}$. $P_{out,r}^{ipSIC}$, $P_{out,r}^{pSIC}$ and ${P_{out,t}}$ is given by \eqref{ipSIC}, \eqref{pSIC} and \eqref{U_m}, respectively.
\end{proposition}

\subsection{Diversity Analysis}\label{Diversity Analysis}
The diversity order is an essential performance metric in wireless networks, which determines the robustness and fading resistance of networks. Specifically, a system with a larger diversity order means that the outage probability decays faster and it is more robust to fading \cite{Laneman2004}, particularly at high SNR. The analysis of diversity order provides a basis for optimizing networks performance and designing more efficient diversity mechanisms. The expression of diversity order is shown as
\begin{align}\label{The definition of diversity order for IRS-NOMA}
{D_{order}} =  - \mathop {\lim }\limits_{P_s^{act} \to \infty } \frac{{\log \left( {P_{out}^{\infty} \left( P_s^{act}  \right)} \right)}}{{\log P_s^{act}  }},
\end{align}
where ${P _{out}^{\infty}  \left( P_s^{act}  \right)}$ denotes the asymptotic expression of outage probability within high SNR region ($P_s^{act} \to \infty $).

The expression for the asymptotic outage probability of $U_r$ with ipSIC can be directly derived from \eqref{ipSIC}, and it is provided by the following corollary.
\begin{corollary}\label{Corollary1:the asymptotic OP of user n with ipSIC under Rician fading channel}
Condition on $P_s^{act} \to \infty $, an asymptotic expression of $U_r$'s outage probability with ipSIC for ASTARS-NOMA is written as

\begin{align}\label{asymptotic ipSIC}
P_{out,r}^{ipSIC,\infty }{\rm{ = }}\sum\limits_{k = 1}^K {\sum\limits_{u = 1}^U {\frac{{\pi {A_k}{\chi _u}}}{{2U\Gamma \left( {{p_r}} \right)}}} } \gamma \left( {{p_r},\sqrt {\frac{{{y_k}{{\hat \gamma }_r}d_s^\alpha \chi _u^\alpha \sigma _{re}^2}}{{\eta _0^2{a_r}{\beta _r}\lambda q_r^2}}} } \right),
\end{align}
where ${\chi _u} = \sqrt {1 - x_u^2} \left( {{x_u}{\rm{ + }}1} \right)$.
\end{corollary}

\begin{remark}\label{Remark1}
As can be observed that the outage probability of $U_r$ with ipSIC is almost constant under the assumptions of $P_s^{act} \to \infty $, i.e., there is an error floor for the outage probability achieved for ${U_r}$ with ipSIC.
By substituting \eqref{asymptotic ipSIC} into \eqref{The definition of diversity order for IRS-NOMA}, the diversity order of $U_r$ with ipSIC can be calculated as $zero$. This is attributed to the effect of residual interference generated by the ipSIC scheme on networks outage performance.
\end{remark}

For $U_r$ with pSIC and $U_t$, the precise outage probability diversity orders can be calculated by employing the Laplace transform.

\begin{corollary}\label{Corollary asymptotic pSIC}
Condition on $P_s^{act} \to \infty $, an asymptotic expression of $U_r$'s outage probability with pSIC for ASTARS-NOMA is given by
\begin{align}\label{The asymptotic pSIC}
P_{out,r}^{pSIC,\infty } = \pi \sum\limits_{u = 1}^U {\frac{{\left( {{x_u}{\rm{ + }}1} \right)\sqrt {1 - x_u^2} }}{{2U\left( {2L} \right)!{\Lambda ^L}}}} {\left[ {_2{F_1}\left( {2,\frac{1}{2};\frac{5}{2};1} \right)} \right]^L} {\left[ {\frac{{{{\hat \gamma }_r}d_s^\alpha }}{{{a_r}P_s^{act}}}\left( {\frac{{\chi _u^\alpha \sigma _0^2}}{{\eta _0^2{\beta _r}\lambda }} + \zeta \frac{{\sigma _s^2}}{{{\eta _0}}}} \right)} \right]^L},
\end{align}
where $\Lambda  = \frac{{3{e^{2\kappa }}}}{{16{{(1 + \kappa )}^2}}}$ and ${}_2{F_1}\left( { \cdot , \cdot ; \cdot ; \cdot } \right)$ stands for the ordinary hypergeometric function \cite[Eq. (9.100)]{2000gradshteyn}.
\begin{proof}
See Appendix~B.
\end{proof}
\end{corollary}

\begin{remark}\label{Remark2}
The $U_r$'s diversity order with pSIC can be calculated as $L$ by substituting  \eqref{The asymptotic pSIC} into \eqref{The definition of diversity order for IRS-NOMA}, which is proportional to the number of ASTARS elements $L$. The $U_r$'s asymptotic outage probability under the pSIC scheme is an oblique line rather than a constant. This indicates that the reflection user $U_r$ with pSIC for ASTARS-NOMA is able to achieve a diversity order of $L$, which is the maximal diversity for the considered scenario.
\end{remark}

\begin{corollary}\label{Corollary3:the asymptotic OP of user m under Rician fading channel}
Condition on $P_s^{act} \to \infty $, an asymptotic expression of $U_t$'s outage probability for ASTARS-NOMA is written as
\begin{align}\label{asymptotic user m}
P_{out,t}^\infty  = \pi \sum\limits_{u = 1}^U {\frac{{\left( {{x_u}{\rm{ + }}1} \right)\sqrt {1 - x_u^2} }}{{2U\left( {2L} \right)!{\Lambda ^L}}}{{\left[ {_2{F_1}\left( {2,\frac{1}{2};\frac{5}{2};1} \right)} \right]}^L}} {\left[ {\frac{{{{\hat \gamma }_t}d_s^\alpha \eta _0^{ - 2}}}{{\left( {{a_t} - {{\hat \gamma }_t}{a_r}} \right)}}\left( {\frac{{y_u^\alpha \sigma _0^2}}{{{\beta _t}\lambda P_s^{act}}} + \zeta \frac{{{\eta _0}\sigma _s^2}}{{P_s^{act}}}} \right)} \right]^L}.
\end{align}
\begin{proof}
The following procedures resemble those in Appendix B.
\end{proof}
\end{corollary}
\begin{remark}\label{Remark3}
The $U_t$'s diversity order is calculated as $L$ by substituting \eqref{asymptotic user m} into \eqref{The definition of diversity order for IRS-NOMA}, which is proportional to the numbers of ASTARS elements. This indicates that $U_t$ of ASTARS-NOMA is also able to achieve the full diversity order.
\end{remark}

\subsection{Delay-limited Transmission}\label{delay-limited mode System throughput}
System throughput in delay-limited transmission situations depends on the outage probability at a target data rate \cite{Zhong2014}. When an outage occurs during data transmission, it means that the data transmission fails and a retransmission is required. At this moment, the system throughput of ASTARS-NOMA with pSIC/ipSIC schemes in delay-limited transmission mode are defined as
\begin{align}\label{The system throughput of delay-limited mode}
{R_{\tau}^{limited}} = \left(1- {P_{out,r}^{\tau}} \right){\hat R_r} + \left( {1 - {P_{out,t}}} \right){\hat R_t},
\end{align}
where $P_{out,r}^{ipSIC}$, $P_{out,r}^{pSIC}$ and ${P_{out,t}}$ are obtained from \eqref{ipSIC}, \eqref{pSIC} and \eqref{U_m}, respectively.
\section{Ergodic Data Rate Analysis}\label{Ergodic Rate}
This section analyzes the ergodic data rate of $U_t$ and $U_r$ with pSIC/ipSIC to reveal the data transmission rate and capability of ASTARS-NOMA networks. The definition of ergodic data rate is shown as
\begin{align}\label{efinition of ergodic rate}
R^{erg} = \mathbb{E}\left[ {{{{\log }_2}} \left( {1 + {\gamma _\varphi }} \right)} \right],
\end{align}
which indicates that the high ergodic data rate achieves high channel capacity. On this basis, we further derive the asymptotic expressions of ergodic data rate for $U_r$ with pSIC/ipSIC and $U_t$. Furthermore, the multiplexing gains for $U_r$ and $U_t$ are discussed in detail.
\subsection{The Ergodic Data Rate of ASTARS-NOMA Networks}
Assuming that $U_r$ can effectively detect information $x_t$ by using the ipSIC method, the expression of $U_r$'s ergodic data rate with ipSIC is then obtained as the following theorem.
\begin{theorem}\label{Theorem ER ipSIC}
Conditioned on the stochastic geometry model and cascade Rician fading channels, a closed-form expression of $U_r$'s ergodic data rate with ipSIC for ASTARS-NOMA is written as
\begin{align}\label{ER ipSIC}
R_{r,ipSIC}^{erg} = \sum\limits_{q = 1}^Q {\sum\limits_{k = 1}^K {\sum\limits_{u = 1}^U {\frac{{\pi \ln \left[ {1 + {{\left( {{x_q}{\vartheta ^{ - 1}}} \right)}^2}} \right]{A_k}{A_q}\left( {{x_u}{\rm{ + }}1} \right)}}{{2U\ln 2\Gamma \left( {{p_r}} \right)x_q^{1 - {p_r}}}}} } },
\end{align}
where $\vartheta  = \frac{1}{{{q_r}}}\sqrt {\frac{{d_s^\alpha }}{{{a_r}P_s^{act}}}\left[ {\zeta \frac{{\sigma _s^2}}{{{\eta _0}}} + \frac{{\chi _u^\alpha }}{{\eta _0^2}}\left( {\frac{{\varepsilon {y_k}P_s^{act}}}{{{\beta _r}\lambda \sigma _{re}^{ - 2}}} + \frac{{\sigma _0^2}}{{{\beta _r}\lambda }}} \right)} \right]} $ and $Q$ is the parameter that guarantee a trade-off between complexity and accuracy of Gauss-Laguerre quadra.
\begin{proof}
See Appendix~C.
\end{proof}
\end{theorem}
\begin{corollary}
For case $\varepsilon =0$, a closed-form expression of $U_r$'s ergodic data rate with pSIC for ASTARS-NOMA is written as

\begin{align}\label{ER pSIC}
R_{r,pSIC}^{erg}=\sum\limits_{q = 1}^Q {\sum\limits_{u = 1}^U {\ln \left( {1 + \frac{{{a_r}{\beta _r}\lambda {{\left( {{x_q}{q_r}{\eta _0}} \right)}^2}P_s^{act}}}{{d_s^\alpha \left( {\chi _u^\alpha \sigma _0^2 + \zeta {\eta _0}{\beta _r}\lambda \sigma _s^2} \right)}}} \right)} } \frac{{\pi {A_q}\left( {{x_u}{\rm{ + }}1} \right)\sqrt {1 - x_u^2} }}{{2U\ln 2\Gamma \left( {{p_r}} \right)x_q^{1 - {p_r}}}}.
\end{align}

\end{corollary}

\begin{theorem}\label{Theorem ER m}
Conditioned on the stochastic geometry model and cascade Rician fading channels, a closed-form expression of $U_t$'s ergodic data rate for ASTARS-NOMA is written as

\begin{align}\label{ER m}
R_t^{erg} = \frac{{\pi {a_t}}}{{2N{a_r}\ln 2}}\sum\limits_{n = 1}^N {\frac{{\sqrt {1 - x_n^2} }}{{1 + {y_n}}}\left\{ {1 - \sum\limits_{u = 1}^U {\frac{{\pi {\chi _u}}}{{2U\Gamma \left( {{p_t}} \right)}}} } \right.} \left. {\gamma \left[ {{p_t},\sqrt {\frac{{{y_n}d_s^\alpha q_t^{ - 2}}}{{P_s^{act}\left( {{a_t} - {y_n}{a_r}} \right)}}\left( {\frac{{y_u^\alpha \sigma _0^2}}{{\eta _0^2{\beta _t}\lambda }} + \zeta \frac{{\sigma _s^2}}{{{\eta _0}}}} \right)} } \right]} \right\},
\end{align}

where ${y_n} = \frac{{\left( {{x_n}{\rm{ + }}1} \right){a_t}}}{{2{a_r}}}$ and ${x_n} = \cos \left( {\frac{{2u - 1}}{{2U}}\pi } \right)$.
\begin{proof}
See Appendix~D.
\end{proof}
\end{theorem}

\subsection{Multiplexing Gains Analysis}\label{Slop Analysis}
We now analyze the multiplexing gains of ASTARS-NOMA networks at high SNRs ($P_s^{act} \to \infty $) to reveal the variation of the ergodic data rate with transmit power \cite{DTse}, which is defined as
\begin{align}\label{The definition of slop}
S =   \mathop {\lim }\limits_{P_s^{act} \to \infty } \frac{{\log \left( {R_{\infty}^{erg}  \left( P_s^{act}  \right)} \right)}}{{\log P_s^{act}  }},
\end{align}
where ${{ {R_{\infty}^{erg}  \left( P_s^{act}  \right)}}}$ is the asymptotic expression of the ergodic data rate within high SNR areas.
\subsubsection{The $U_r$'s multiplexing gain with ipSIC}
Based on \eqref{ER ipSIC}, when $P_s^{act} \to \infty $ the asymptotic expression of $U_r$'s ergodic data rate with ipSIC for ASTARS-NOMA is written as
\begin{align}\label{asymptotic ER ipSIC}
R_{r,ipSIC}^{erg,\infty } =\sum\limits_{q = 1}^Q {\sum\limits_{k = 1}^K {\sum\limits_{u = 1}^U {\ln \left[ {1 + {{\left( {{\eta _0}{x_q}{q_r}} \right)}^2}\frac{{\lambda {\beta _r}{a_r}}}{{d_s^\alpha \chi _u^\alpha {y_k}\sigma _{re}^2}}} \right]} } }\frac{{\pi {A_k}{A_q}\left( {{x_u}{\rm{ + }}1} \right)}}{{2U\ln 2\Gamma \left( {{p_r}} \right)}}x_q^{{p_r} - 1}.
\end{align}
\begin{remark}\label{remark ER ipSIC}
By substituting \eqref{asymptotic ER ipSIC} into \eqref{The definition of slop}, a multiplexing gain of $zero$ for $U_r$ with ipSIC is obtained. It implies that even when the $P_s^{act}$ is sufficiently high, the ergodic data rate will not increase.
\end{remark}
\subsubsection{The $U_r$'s multiplexing gain with pSIC}
We obtain an upper bound on the ergodic data rate by invoking Jensen's inequality, which is written as
\begin{align}
R_{r,pSIC}^{erg} = \mathbb{E}\left[ {{{\log }_2} \left( {1 + {\gamma _r}} \right)} \right] \le {{\log }_2} \left[ {1 + \mathbb{E}\left( {{\gamma _r}} \right)} \right].
\end{align}

Based on the above inequality, when $P_s^{act} \to \infty $ an upper bound expression of $U_r$'s ergodic data rate with pSIC for ASTARS-NOMA networks is written as
}
\begin{align}\label{asymptotic ER pSIC}
R_{r,erg}^{bound} = \log \left\{ {1 + \frac{{\Xi L\lambda P_s^{act}\left[ {D\left( {X_r^l} \right) + L{E^2}\left( {X_r^l} \right)} \right]}}{{d_s^\alpha \left[ {\lambda {\beta _r}{\eta _0}\sigma _s^2\zeta \left( {2 + \alpha } \right) + 2{D^{2 + \alpha }}\sigma _0^2} \right]}}} \right\},
\end{align}

where $\Xi  = \eta _0^2{a_r}{\beta _r}\left( {2 + \alpha } \right)$.

\begin{remark}\label{remark ER pSIC}
The multiplexing gain for $U_r$ with pSIC is equal to $one$ by inserting  \eqref{asymptotic ER pSIC} into \eqref{The definition of slop}, which is due to the limitation of power allocation factors ratio at high SNRs.
\end{remark}

\subsubsection{The $U_t$'s multiplexing gain}
When $P_s^{act} \to \infty $ the asymptotic ergodic data rate expression of $U_t$ for ASTARS-NOMA networks can be obtained directly from \eqref{ER m} as
\begin{align}\label{asymptotic ER m}
R_t^{erg,\infty } = \frac{{\pi {a_t}}}{{2N{a_r}\ln 2}}\sum\limits_{n = 1}^N {\frac{{\sqrt {1 - x_n^2} }}{{1 + {y_n}}}} .
\end{align}
\begin{remark}\label{remark ER m}
The multiplexing gain for $U_t$ is equal to $zero$, by inserting \eqref{asymptotic ER m} into \eqref{The definition of slop}, which has the same conclusion as in \bf{Remark \ref{remark ER ipSIC}}.
\end{remark}

\subsection{Delay-tolerant Transmission}\label{delay-tolerated mode System throughput}
The data transmission is capped at the ergodic data rate in delay-tolerant transmission situations because the codeword can experience all channel realizations \cite{Zhong2014}. In this scenario, the system throughput of ASTARS-NOMA with pSIC/ipSIC schemes in delay-tolerated transmission mode are defined as
\begin{align}\label{The system throughput of delay-tolerant mode}
{R_{\vartheta}^{tolerant }} = R_{r,\tau}^{erg}  + {R_{t}^{erg}},
\end{align}
where $R_{r,ipSIC}^{erg}$, $R_{r,pSIC}^{erg}$ and $R_{t}^{erg}$ can be obtained from \eqref{ER ipSIC}, \eqref{ER pSIC} and  \eqref{ER m}.

\section{Simulation Results}\label{Numerical Results}
\begin{table}[t!]
\centering
\caption{Simulation Parameters for ASTARS-NOMA networks.}
\tabcolsep3pt
\renewcommand\arraystretch{0.8} 
\begin{tabular}{|l|l|}
\hline
Monte Carlo simulations repeated  &  ${10^6}$ iterations \\
\hline
Rician factor   &  $\kappa=-5$ dB   \\
\hline
Amplification factor &  $\lambda  = 5$   \\
\hline
Number of ASTARS elements  &  $L = 10$   \\
\hline
\multirow{1}{*}{Coverage radius of ASTARS}& \multirow{1}{*}{$D=35$ m } \\
\hline
Distance from BS to ASTARS  &  $d_{s}=50$ m   \\
\hline
\multirow{1}{*}{Amplitude coefficients of ASTARS elements} &  \multirow{1}{*}{$\beta_r=0.7$, $\beta_t=0.3$}   \\
\hline
\multirow{1}{*}{The power allocation factors of $U_r$ and $U_t$}&  \multirow{1}{*}{$a_r=0.3$, $a_t=0.7$}   \\
\hline
\multirow{2}{*}{Noise power} &  \multirow{1}{*}{$\sigma _{s}^2 = - 70 $ dBm}  \\
 & \multirow{1}{*}{$\sigma _{0}^2 = \sigma _{re}^2 =  - 90 $ dBm} \\
\hline
\multirow{2}{*}{Pass loss factors}   &  \multirow{1}{*}{$\alpha  = 2$}  \\
& \multirow{1}{*}{${\eta _0}=-30$ dB } \\
\hline
\multirow{2}{*}{Target data rates for $U_r$ and $U_t$ } & \multirow{1}{*}{${{\hat R}_r}=1$ BPCU}  \\
                                                & \multirow{1}{*}{${{\hat R}_t}=1$ BPCU} \\
\hline
\cline{1-2}
\end{tabular}
\label{parameter}
\end{table}
This section provide the computer simulation results to verify the correctness of the theoretical formulas in Sections III and IV. The simulation settings used, unless otherwise specified, are displayed in Table I. The complexity-accuracy trade-off parameters $K$, $Q$ and $U$ are set to $10^3$. To highlight the performance of ASTARS-NOMA networks, the ARIS-NOMA, ASTARS-OMA and PSTARS-NOMA networks are selected as benchmarks. In particular,  the total power budgets of ASTARS and PSTARS-aided networks  are respectively given by $Q_{tot}^{act} = P_s^{act} + P_r^{act} + L\left( {{P_c} + {P_d}} \right)$ and $Q_{tot}^{pas} = P_s^{pas} + L{P_c}$ \cite{LongARIS} \cite{PanARIS}, where $P_r^{act}$ is the output signal power of ASTARS/ARIS, the power used by the phase shift control circuit in each active element is ${P_c} =  - 20$ dBm and the direct current bias power used by the amplifier placed in each ASTARS/ARIS element is indicated as ${P_d} =  - 20$ dBm. To achieve fair comparisons, $Q_{tot}^{act} $ is set to be the same as $Q_{tot}^{pas} $. Moreover, in ARIS-NOMA networks, to obtain ${360^ \circ }$ coverage, we employ a surface made up of one transmit-only RIS and one reflect-only RIS \cite{STARXu}. It's also important to note that ASTARS-OMA networks employ time division multiple access, which takes twice as long as NOMA for serving the two users.
\subsection{Outage Probability}

\begin{figure}[htbp]
\centering
\subfigure[Outage probability versus system power budget ${Q_{tot}}$.]{\label{diff_user}
\begin{minipage}[t]{0.5\linewidth} 
\centering
\includegraphics[width=0.9\textwidth,height=0.7\textwidth]{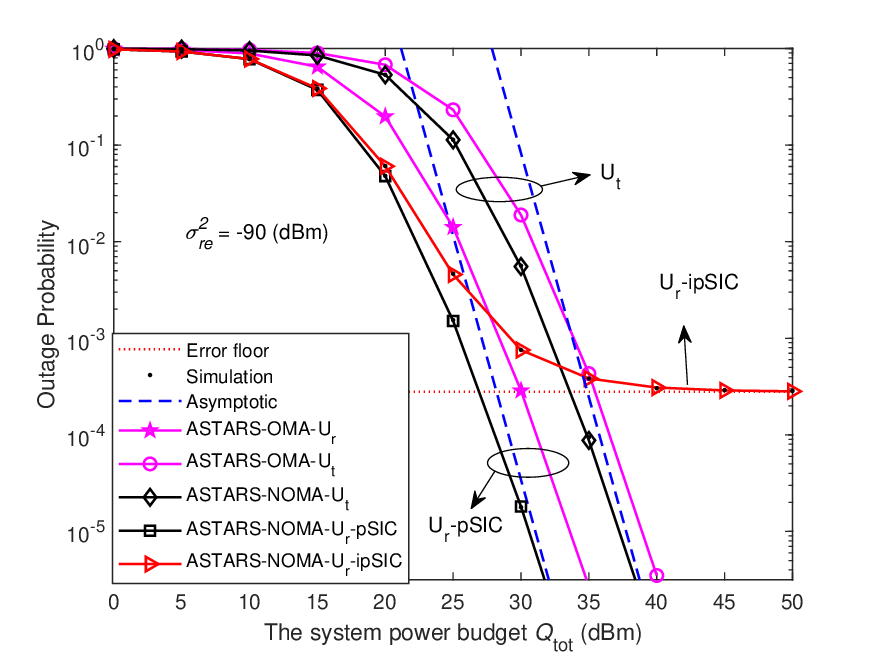}
\end{minipage}%
}%
\subfigure[Outage probability versus BS's transmit power $P_s^{act}$.]{\label{Diff_K}
\begin{minipage}[t]{0.5\linewidth} 
\centering
\includegraphics[width=0.9\textwidth,height=0.7\textwidth]{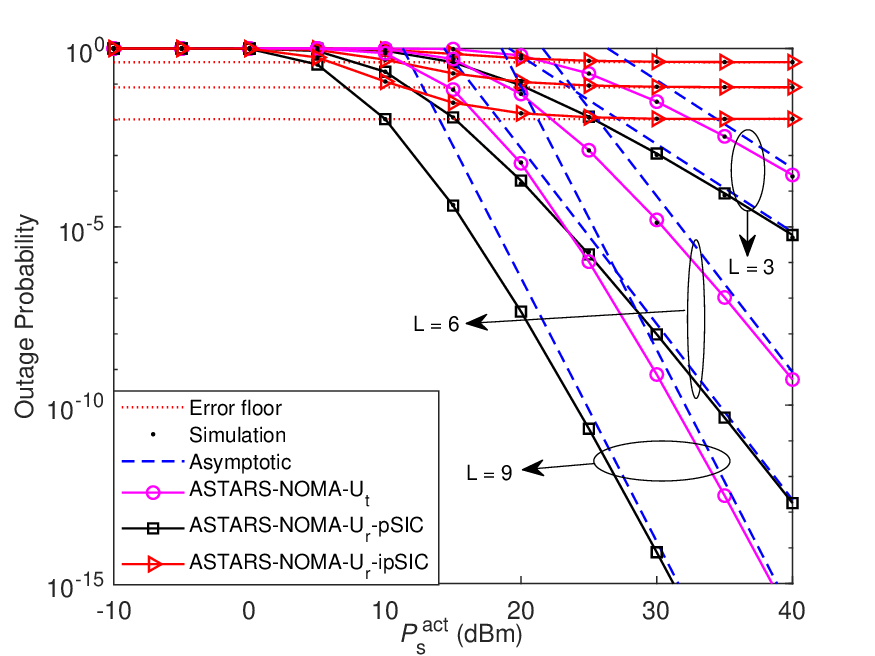}
\end{minipage}%
}%
\centering
\caption{Outage probability of ASTARS-NOMA networks.}
\end{figure}

In Fig. \ref{diff_user}, we plot the outage probability of $U_r$ and $U_t$ versus the system power budget. The outage probability curves of $U_t$ and $U_r$ with pSIC/ipSIC curves are plotted by \eqref{ipSIC}, \eqref{pSIC} and \eqref{U_m}, respectively. This figure demonstrates that the obtained analytical expressions match the simulation results exactly, which validates the accuracy of the analytical methods applied. The blue dotted lines for asymptotic outage probability are ploted based on \eqref{asymptotic ipSIC}, \eqref{The asymptotic pSIC} and \eqref{asymptotic user m}, respectively. They perfectly match the outage probability curves of ASTARS-NOMA within high SNR region, proving that our asymptotic approach is accurate. One phenomenon is that the outage performance of $U_r$ with pSIC and $U_t$ for ASTARS-NOMA outperforms that of ASTARS-OMA, which is due to the following two reasons. 1) NOMA is able to achieve better fairness in outage performance between paring users; and 2) The performance of ASTAR-NOMA can be further enhanced by the better compatibility between ASTARS and NOMA. Another phenomenon is that, after ${Q_{tot}}$ exceeds $30$ dBm, the outage performance of $U_r$ with ipSIC for ASTARS-NOMA is worse than that of ASTARS-OMA. In high SNR area, it converges to an error floor. This is attributed that the $U_r$ in ASTARS-NOMA networks suffers from residual interference caused by ipSIC, which confirms the conclusions made in \textbf{Remark \ref{Remark1}}.
Moreover, Fig. \ref{Diff_K} displays the outage probability of $U_r$ and $U_t$ for ASTARS-NOMA networks versus transmit power of BS under $L = \left\{ {3,6,9} \right\}$. The figure shows that as the number of ASTARS elements rises, the outage probability falls and its slope rises. This phenomenon is caused by the fact that the diversity orders of $U_t$ and $U_r$ with pSIC are proportional to the number of ASTARS elements, which confirms the conclusions made in \textbf{Remark \ref{Remark2}} and \textbf{Remark \ref{Remark3}}.

\begin{figure}[htbp]
\centering
\subfigure[Outage probability versus number of ASTARS/ARIS elements $L$, with $Q_{tot}^{act} =Q_{tot}^{pas}= 20$ dBm, $\lambda  = 10$ and $\sigma _s^2 =  - 30$ dBm.]{\label{Diff_L}
\begin{minipage}[t]{0.5\linewidth} 
\centering
\includegraphics[width=0.9\textwidth,height=0.7\textwidth]{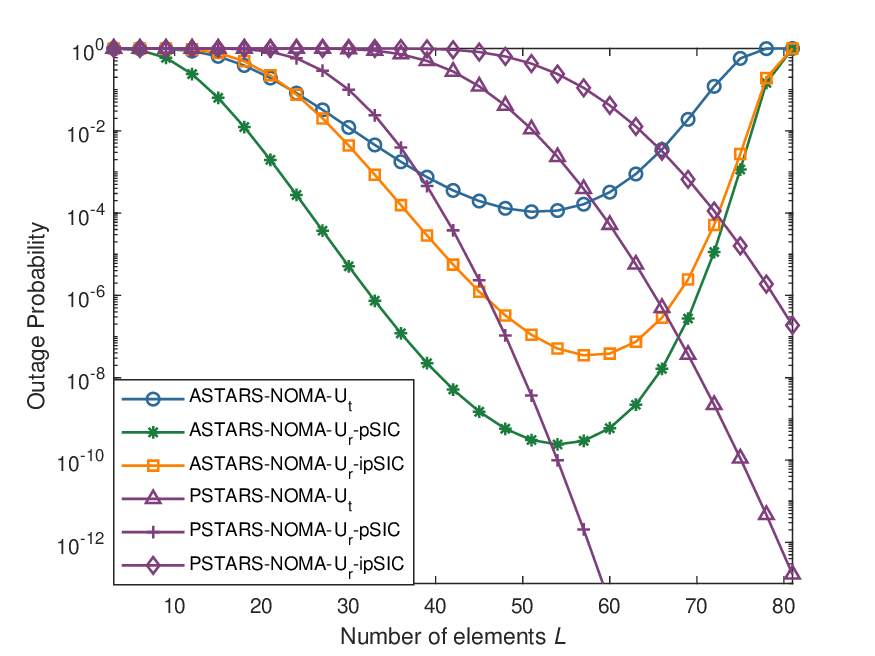}
\end{minipage}%
}%
\subfigure[System outage probability versus system power budget ${Q_{tot}}$.]{\label{Baseline}
\begin{minipage}[t]{0.5\linewidth} 
\centering
\includegraphics[width=0.9\textwidth,height=0.7\textwidth]{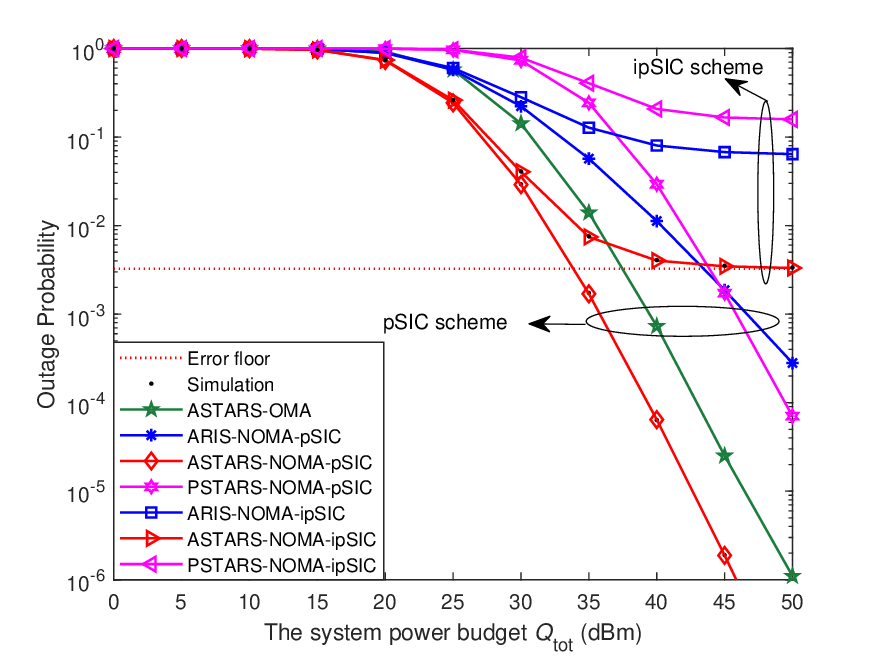}
\end{minipage}%
}%
\centering
\caption{Outage probability of ASTARS-NOMA networks.}
\end{figure}

In Fig. \ref{Diff_L}, we plot the outage probability of ASTARS-NOMA versus number of ASTARS elements $L$, with $Q_{tot}^{act} =Q_{tot}^{pas}= 20$ dBm, $\lambda  = 10$ and $\sigma _s^2 =  - 30$ dBm. As can be observed that with ASTARS elements increase, the outage probability first decreases and then gradually increases. This trend can be attributed to the complex interactions between various factors. On the one hand, the introduction of more ASTARS elements can enhance the spatial degrees of freedom and thus reduce the outage probability. This is because that the increased degrees of freedom enable a more efficient use of spatial domain, resulting in better signal quality and stronger channel gains. On the other hand, using too many ASTARS elements can lead to a large amount of thermal noise, which severely hinders the user's ability to decode the signal. This counteracts the channel gains generated by spatial degrees of freedom and leads to a surge in outage probability. Hence, the optimization of ASTARS-NOMA networks is a balance between the numbers of active component and the spatial degrees of freedom, with the ultimate goal of minimizing outage probability.

Fig. \ref{Baseline} displays the system outage probability of ASTARS-NOMA and different benchmarks versus the system power budget. As can be shown that the outage performance of ASTARS-NOMA perform better than the PSTARS-NOMA. This can be interpreted that the ASTARS elements allocate a portion of power budget to amplify the input radio signals, which enhances the received SNR at paring users. This also confirms that ASTARS is an effective technology for combating multiplicative fading loss. One occurrence is that ASTARS-NOMA achieves better outage performance than ARIS-NOMA. In addition, the outage curves of ASTARS-NOMA are steeper than the OMA ones. This can be explained by the fact that ASTARS is able to achieve a better diversity order than ARIS.

\begin{figure}[htbp]
\centering
\subfigure[System outage probability versus reflection amplitude coefficient and power allocation factor, with $P_s^{act} = 20$ dBm.]{\label{Diff_betar}
\begin{minipage}[t]{0.5\linewidth} 
\centering
\includegraphics[width=0.9\textwidth,height=0.7\textwidth]{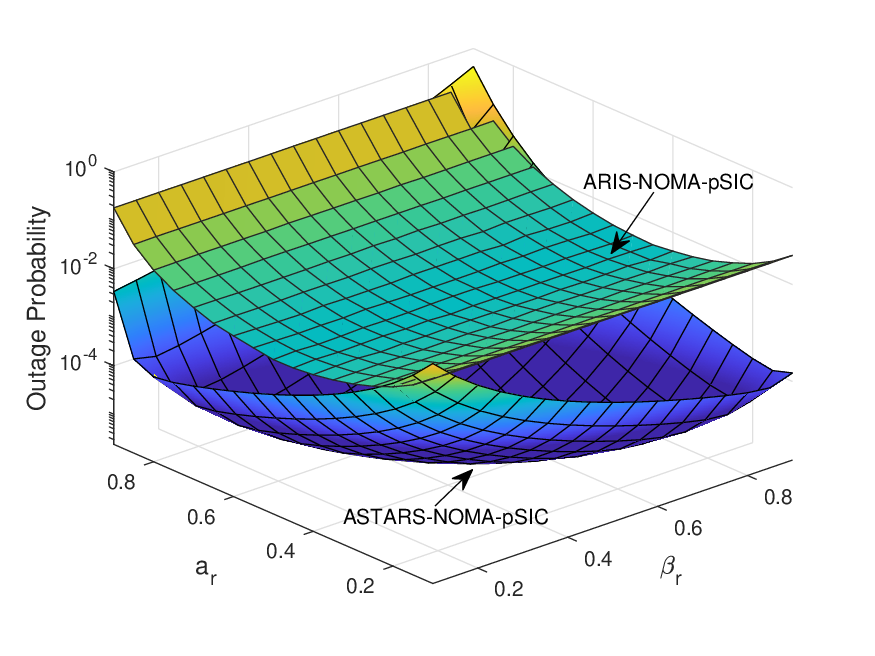}
\end{minipage}%
}%
\subfigure[Outage probability versus amplification factor $\lambda$, with $P_s^{act} = 25$ dBm and $\sigma _s^2 =  - 50$ dBm.]{\label{Diff_lama}
\begin{minipage}[t]{0.5\linewidth} 
\centering
\includegraphics[width=0.9\textwidth,height=0.7\textwidth]{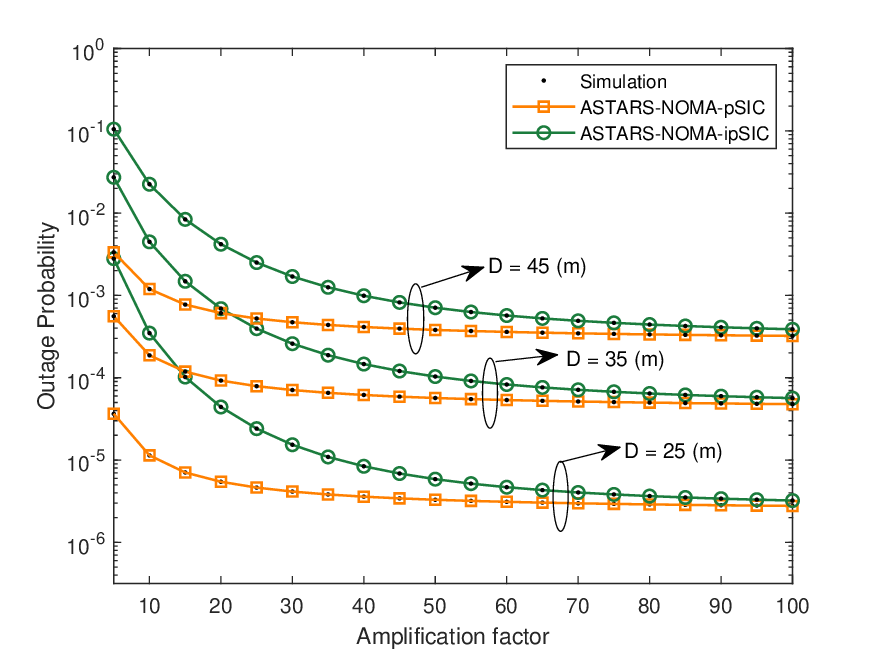}
\end{minipage}%
}%
\centering
\caption{Outage probability of ASTARS-NOMA networks.}
\end{figure}

In Fig. \ref{Diff_betar}, we show the system outage probability of ASTARS-NOMA versus reflection amplitude coefficient and power allocation factor under $Q_{act} = 20$ dBm. The curved surface for outage probability of ASTARS-NOMA is plotted according to \eqref{system outage}. Due to the effects of both amplitude coefficients and power allocation factors, the curved surface for outage probability of ASTARS-NOMA takes on a valley-like shape. This suggests the existence of a solution that satisfies the minimum system outage probability for the given parameter. Additionally, it can be shown that by adjusting $a_r$, the ARIS-NOMA's outage probability can be reduced. However, in most cases of parameter setting, the outage performance of ARIS-NOMA is inferior to that of ASTARS-NOMA. This is because that ASTARS can introduce more spatial degrees of freedom for NOMA networks, which enhances the outage performance. It is worth noting that when with some extreme parameter choices, such as $\beta_r=0.1$ and $a_r=0.1$, the performance of ASTARS-NOMA networks may be even worse than that of ARIS-NOMA networks. This indicates that optimizing the amplitude coefficients and power allocation factors is essential to reducing the outage probability of ASTARS-NOMA.

In Fig. \ref{Diff_lama}, we plot the system outage probability of ASTARS-NOMA versus amplification factor $\lambda$, with $P_s^{act} = 25$ dBm and $\sigma _s^2 =  - 50$ dBm. One phenomenon is that the outage probability of ASTARS-NOMA first reduces dramatically as the amplification factor steadily rises and then tends to stabilize. This is due to the fact that the larger amplification factors helps to improve the users' received SNR, thus enhancing the outage performance. However, while increasing the received signal strength at the users, also introduces a large amount of thermal noise, which interferes with the decoding of user signals. As the amplification factor increases, a balance is achieved between the gain of the enhanced signal and the loss of the enhanced noise such that the outage probability remains constant. Another observation is that reducing the deployment range of users improves outage performance. This is because that a smaller deployment range diminishes the effect of path loss on the ASTARS-NOMA's outage probability.

\subsection{Ergodic Data Rate}
\begin{figure}[htbp]
\centering
\subfigure[Ergodic data rate versus system power budget ${Q_{tot}}$ with $a_r=0.2$, $a_t=0.8$.]{\label{ER1}
\begin{minipage}[t]{0.5\linewidth} 
\centering
\includegraphics[width=0.9\textwidth,height=0.7\textwidth]{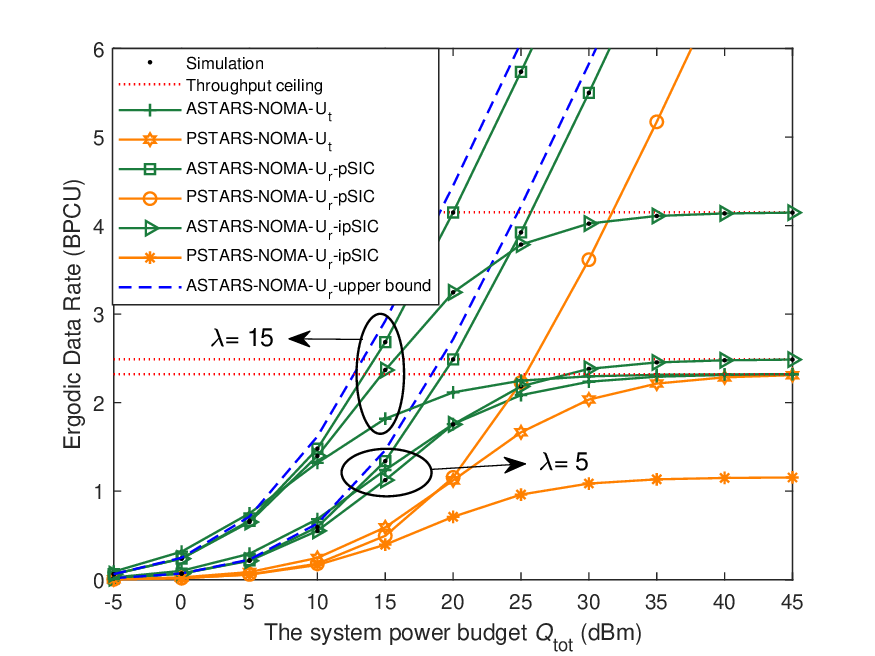}
\end{minipage}%
}%
\subfigure[Ergodic data rate versus system power budget ${Q_{tot}}$ with $a_r=0.2$, $a_t=0.8$.]{\label{ER2}
\begin{minipage}[t]{0.5\linewidth} 
\centering
\includegraphics[width=0.9\textwidth,height=0.7\textwidth]{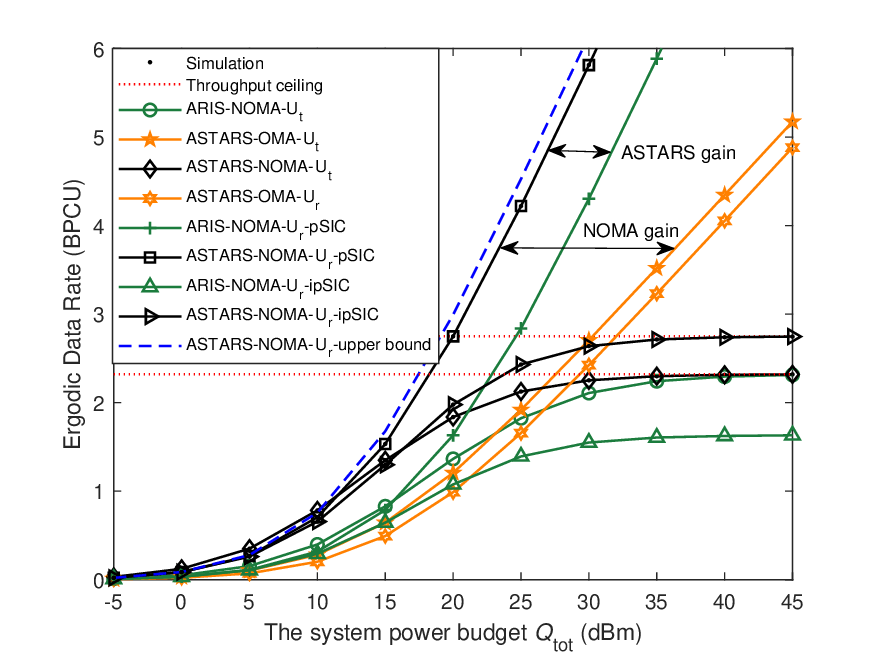}
\end{minipage}%
}%
\centering
\caption{Ergodic data rate of ASTARS-NOMA networks.}
\end{figure}

In Fig. \ref{ER1}, we present ergodic data rate of ASTARS-NOMA and PSTARS-NOMA versus system power budget with $a_r=0.2$, $a_t=0.8$.
The ergodic data rate curves for ASTARS-NOMA networks are drawn from \eqref{ER ipSIC}, \eqref{ER pSIC} and \eqref{ER m}, respectively. According to \eqref{asymptotic ER ipSIC}, \eqref{asymptotic ER pSIC} and \eqref{asymptotic ER m}, the asymptotic ergodic data rates are illustrated. This figure indicates that the $U_t$'s ergodic data rate converges towards the upper limit of the throughput, resulting in $zero$ high SNR slope. The ergodic data rate of ipSIC stops increasing with a rise in transmit power at high SNRs due to the effects of residual interference, which in accordance with the discussion in \textbf{Remark \ref{remark ER ipSIC}}. One phenomenon is that the $U_r$'s ergodic data rates with pSIC/ipSIC of ASTARS-NOMA are higher than those of PSTARS-NOMA. This is due to the fact that ASTARS is able to increase the strength of users' received signals, which further increase the average data transmission rate of networks over an extended period. Another phenomenon is that the ergodic data rates of ASTARS-NOMA with a larger power amplification factor is more efficient. This suggests that increasing the power amplification factor can enhance the receiving SNR and thus improve the ergodic data rate of ASTARS-NOMA.

Fig. \ref{ER2} compares the ergodic data rates of ASTARS-NOMA with ARIS-NOMA and ASTARS-OMA. This figure indicates that the $U_r$'s ergodic data rate with pSIC/ipSIC of ASTARS-NOMA outperform that of ARIS-NOMA and ASTARS-OMA. For the OMA transmission, it takes twice as long to serve two users as NOMA transmission. As a result, the slope of OMA transmission is only half of that of NOMA transmission, which is the reason for its lower ergodic data rate. The reason why ARIS-NOMA networks have lower ergodic data rate compared to ASTARS-NOMA networks is that they cannot provide the same spatial degrees of freedom as ASTARS networks do. Another phenomenon is that the ergodic data rate of $U_t$ for ASTARS-NOMA outperforms ARIS-NOMA at low SNRs, while they reach the same upper limit of rate within high SNR region. This can be explained by using the conclusion of \textbf{Remark \ref{remark ER m}} that the upper limit of ${U_t}$'s ergodic data rate in the NOMA network is related to the power allocation factors.

\begin{figure}[t!]
    \begin{center}
        \includegraphics[width=2.9in,  height=2.2in]{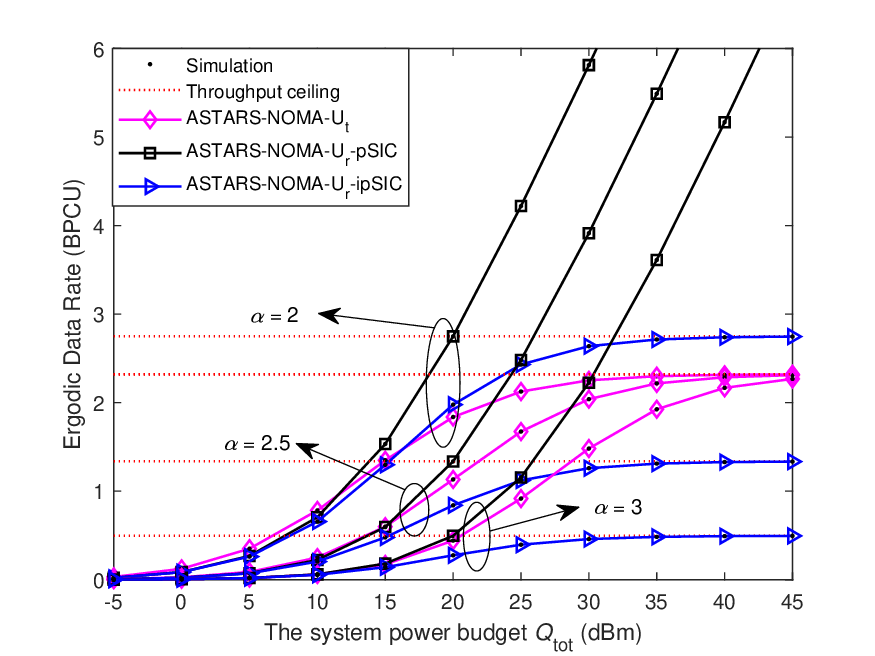}
        \caption{Ergodic data rate versus system power budget ${Q_{tot}}$ with $a_r=0.2$, $a_t=0.8$.}
        \label{ER3}
    \end{center}
\end{figure}

Additionally, Fig. \ref{ER3} plots the ergodic data rate of ASTARS-NOMA networks versus system power budget with different path loss exponents, i.e., $\alpha$. In ASTARS-NOMA networks, $\alpha$ is a parameter to describe the signal power attenuation as it propagates through the wireless channels. A few real-world channel models are to be adopted, depending on the choice of $\alpha$. For example, $\alpha  = 2$ denotes the free space propagation case, $\alpha  = 2.5$ denotes the scenario with obstacles and $\alpha  = 3$ denotes the urban cellular networks. One can observe that as $\alpha$ increases, the channel conditions of ASTARS-NOMA networks become progressively worse, resulting in a deterioration of outage performance. This indicates that a reasonable $\alpha$ should be selected when studying the ASTARS-NOMA networks in different practical scenarios.

\subsection{System Throughput}

\begin{figure}[htbp]
\centering
\subfigure[Delay-tolerant transmission.]{\label{Throughput}
\begin{minipage}[t]{0.5\linewidth} 
\centering
\includegraphics[width=0.9\textwidth,height=0.7\textwidth]{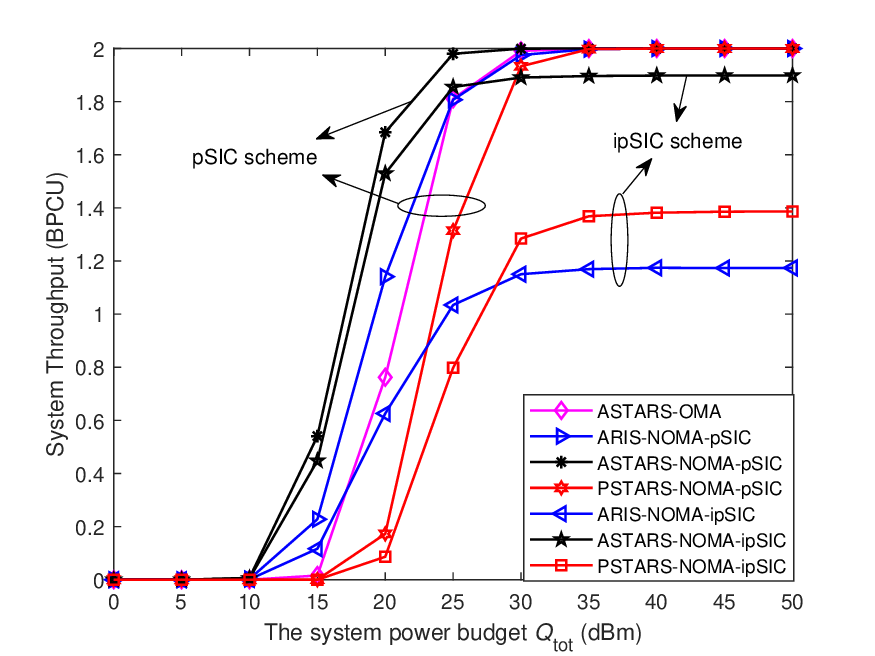}
\end{minipage}%
}%
\subfigure[Delay-limited transmission.]{\label{Throughput_l}
\begin{minipage}[t]{0.5\linewidth} 
\centering
\includegraphics[width=0.9\textwidth,height=0.7\textwidth]{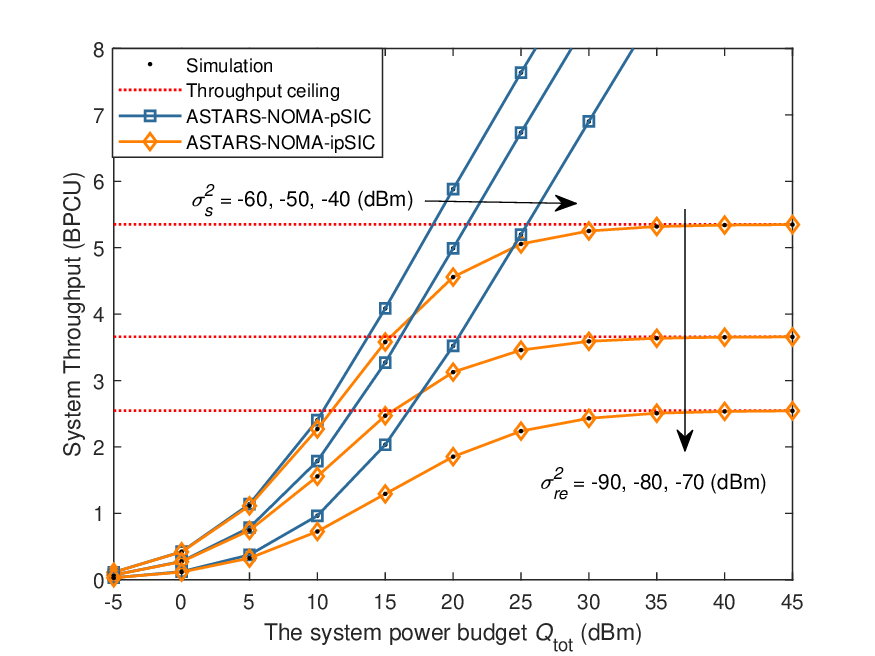}
\end{minipage}%
}%
\centering
\caption{System throughput versus system power budget ${Q_{tot}}$.}
\end{figure}

In Fig. \ref{Throughput}, we present the system throughput of ASTARS-NOMA with pSIC/ipSIC versus system power budget in the delay-limited transmission mode. According to \eqref{The system throughput of delay-limited mode}, the system throughput curves of ASTARS-NOMA with pSIC/ipSIC are drawn. One phenomenon is that the system throughput of ASTARS-NOMA outperform other comparison baselines under pSIC scheme. This is attributed to the fact that outage probability under delay-limited transmission model determines the throughput of ASTARS-NOMA networks. It can also be seen that the NOMA networks with ipSIC fail to reach the target rate even when the transmit power is large. This is due to the fact that residual interference limits the performance gains from increasing transmit power at high SNRs.

Fig. \ref{Throughput_l} shows the system throughput of ASTARS-NOMA versus system power budget in the delay-tolerant transmission model, with $a_r=0.2$, $a_t=0.8$ and $\alpha=2.3$. According to \eqref{The system throughput of delay-tolerant mode}, the system throughput curves of ASTARS-NOMA with pSIC/ipSIC schemes are shown. One phenomenon is that the system throughput ceiling of ASTARS-NOMA under ipSIC scheme increases as residual interference strength diminishes. Another phenomenon is that reducing the noise intensity generated by active devices can improve the system throughput of ASTARS-NOMA with pSIC. This indicates that the design of low-power and low-interference hardware architecture is essential to improve the performance of ASTARS-NOMA.

\section{Conclusion}\label{Conclusion1}
In this article, we have studied the novel ASTARS-NOMA networks with randomly deployed users, which can mitigate the multiplicative fading loss and achieve full-space smart radio environments. Specifically, we have obtained analytical expressions of outage probability and ergodic data rate for ASTARS-NOMA networks with pSIC/ipSIC scheme. To gain further insights, the asymptotic expressions of outage probability and ergodic data rate were also obtained within high SNR region. On this basis, we have analyzed the diversity orders and multiplexing gains for paring users. Simulation results demonstrated that the performance of ASTARS-NOMA outperforms the PSTARS-NOMA, ARIS-NOMA and ASTARS-OMA for the same power consumption. For hardware configuration, the ASTARS-NOMA networks require an appropriate power amplification factors and number of ASTARS elements to ensure that the thermal noise interference is within a reasonable range.

\appendices
\section*{Appendix~A: Proof of Theorem \ref{Theorem1:the OP of user n with ipSIC under Rician fading channel}}\label{Appendix:A}
\renewcommand{\theequation}{A.\arabic{equation}}
\setcounter{equation}{0}

The top of the next page displays the expression of $U_r$'s outage probability with ipSIC by substituting \eqref{SINR nm} and \eqref{SINR n} into \eqref{OP event n}. After combining the probability events, \eqref{the total OP expression of user n with ipSIC} can be simplified to
\begin{figure*}[!t]
\begin{align}\label{the total OP expression of user n with ipSIC}
&P_{out,r}^{ipSIC} ={\rm{ Pr}}\left[ {\frac{{{a_t}\lambda {\beta _r}P_s^{act}{{\left| {{\bf{h}}_r^H{{\bf{\Phi }}_r}{{\bf{h}}_s}} \right|}^2}}}{{{a_r}\lambda {\beta _r}P_s^{act}{{\left| {{\bf{h}}_r^H{{\bf{\Phi }}_r}{{\bf{h}}_s}} \right|}^2} + \lambda {\beta _r}{{\left| {{\bf{h}}_r^H{{\bf{\Phi }}_r}{{\bf{n}}_s}} \right|}^2} + \sigma _0^2}} \le {{\hat \gamma }_t}} \right] \nonumber \\
&+\left[ {\frac{{{a_t}\lambda {\beta _r}P_s^{act}{{\left| {{\bf{h}}_r^H{{\bf{\Phi }}_r}{{\bf{h}}_s}} \right|}^2}}}{{{a_r}\lambda {\beta _r}P_s^{act}{{\left| {{\bf{h}}_r^H{{\bf{\Phi }}_r}{{\bf{h}}_s}} \right|}^2} + \lambda {\beta _r}{{\left| {{\bf{h}}_r^H{{\bf{\Phi }}_r}{{\bf{n}}_s}} \right|}^2} + \sigma _0^2}} > {{\hat \gamma }_t},\frac{{{a_r}\lambda {\beta _r}P_s^{act}{{\left| {{\bf{h}}_r^H{{\bf{\Phi }}_r}{{\bf{h}}_s}} \right|}^2}}}{{\lambda {\beta _r}{{\left| {{\bf{h}}_r^H{{\bf{\Phi }}_r}{{\bf{n}}_s}} \right|}^2} + \varepsilon {{\left| {{h_{re}}} \right|}^2}P_s^{act} + \sigma _0^2}} \le {{\hat \gamma }_r}} \right].
\end{align}
\hrulefill \vspace*{0pt}
\end{figure*}
\begin{align}\label{A2}
P_{out,r}^{ipSIC} = {\rm{Pr}}\left[ {{{\left| {{\bf{h}}_r^H{{\bf{\Phi }}_r}{{\bf{h}}_s}} \right|}^2}} \right. \le \left. {\frac{{{{\hat \gamma }_r}}}{{{a_r}P_s^{act}}}\left( {{{\left| {{\bf{h}}_r^H{{\bf{\Phi }}_r}{{\bf{n}}_s}} \right|}^2} + \frac{{\varepsilon {{\left| {{h_{re}}} \right|}^2}}}{{{\beta _r}\lambda }}P_s^{act} + \frac{{\sigma _0^2}}{{{\beta _r}\lambda }}} \right)} \right].
\end{align}

With help of the topics discussed in Section \ref{Thermal noise}, $U_r$'s outage probability with ipSIC can be further calculated as
\begin{align}\label{A3}
P_{out,r}^{ipSIC}{\rm{ = Pr}}\left\{ {\underbrace {{{\left| {\sum\nolimits_{l = 1}^L {\left| {h_s^lh_r^l} \right|} } \right|}^2}}_{{X_r}}} \right.\left. { \le \frac{{{{\hat \gamma }_r}d_s^\alpha }}{{{a_r}\eta _0^2}}\left[ {\zeta \frac{{{\eta _0}\sigma _s^2}}{{P_s^{act}}} + \underbrace {d_r^\alpha }_Z\left( {\frac{\varepsilon }{{{\beta _r}\lambda }}\underbrace {{{\left| {{h_{re}}} \right|}^2}}_Y + \frac{{\sigma _0^2}}{{{\beta _r}\lambda P_s^{act}}}} \right)} \right]} \right\},
\end{align}
where $\zeta  = L\left( {\frac{{L\kappa  + 1}}{{\kappa  + 1}}} \right)$.

The PDF of $Y$ can be express as ${f_Y}\left( y \right) = \frac{1}{{\sigma _{re}^2}}{e^{ - \frac{y}{{\sigma _{re}^2}}}}$, and $X_r$'s CDF and $Z$'s PDF denoted by \eqref{XCDF} and \eqref{dnPDF}, respectively. Combining \eqref{XCDF}, \eqref{dnPDF} and PDF of $Y$, \eqref{A3} can be converted into integral form as
\begin{small}
\begin{align}
P_{out,r}^{ipSIC}{\rm{ = }}\int_0^\infty  {\int_0^D {\frac{1}{{\sigma _{re}^2}}{e^{ - \frac{y}{{\sigma _{re}^2}}}}\frac{{2z}}{{{D^2}\Gamma \left( {{p_r}} \right)}}} }  \gamma \left\{ {{p_r},\frac{1}{{{q_r}}}\sqrt {\frac{{{{\hat \gamma }_r}d_s^\alpha }}{{{a_r}P_s^{act}}}\left[ {\zeta \frac{{\sigma _s^2}}{{{\eta _0}}} + \frac{{{z^\alpha }}}{{\eta _0^2}}\left( {\frac{{\varepsilon yP_s^{act}}}{{{\beta _r}\lambda }} + \frac{{\sigma _0^2}}{{{\beta _r}\lambda }}} \right)} \right]} } \right\}dzdy,
\end{align}
\end{small}

By applying Gauss-Chebyshev quadrature \cite[Eq. (8.8.4)]{Hildebrand1987introduction}, the definite integral of above expression can be calculated as
\begin{small}
\begin{align}
P_{out,r}^{ipSIC}{\rm{ = }}\frac{\pi }{{2U}}\sum\limits_{u = 1}^U {\int_0^\infty  {{e^{ - \frac{y}{{\sigma _{re}^2}}}}} \frac{{\left( {{x_u}{\rm{ + }}1} \right)}}{{\sigma _{re}^2\Gamma \left( {{p_r}} \right)}}\sqrt {1 - x_u^2} }  \gamma \left\{ {{p_r},\frac{1}{{{q_r}}}\sqrt {\frac{{{{\hat \gamma }_r}d_s^\alpha }}{{{a_r}P_s^{act}}}\left[ {\zeta \frac{{\sigma _s^2}}{{{\eta _0}}} + \frac{{\chi _u^\alpha }}{{\eta _0^2}}\left( {\frac{{\varepsilon yP_s^{act}}}{{{\beta _r}\lambda }} + \frac{{\sigma _0^2}}{{{\beta _r}\lambda }}} \right)} \right]} } \right\}dy.
\end{align}
\end{small}

It can be seen that the above equation contains the term ${e^{ - at}}$ and the limits of integration are $0$ to infinity. This type of integral equation can be calculated by applying Gauss-Laguerre quadrature formula \cite[Eq. (8.6.5)]{Hildebrand1987introduction}, i.e., $\int_0^\infty  {{e^{ - at}}f\left( t \right)dt}  = \frac{{{1}}}{a}\sum\limits_{k = 1}^K {{A_k}f\left( {{{\frac{{{x_k}}}{a}}}} \right)} $. After some algebraic manipulations, we can obtain \eqref{ipSIC}. The proof is complete.

\section*{Appendix~B: Proof of Corollary \ref{Corollary asymptotic pSIC}}\label{Appendix:B}
\renewcommand{\theequation}{B.\arabic{equation}}
\setcounter{equation}{0}
To obtain the accurate asymptotic outage probability, the Laplace transform is applied in the following proof process. The Laplace
transform formula of the PDF for ${X_\varphi ^l}$ is calculated by \cite[Eq. (6.621.3)]{2000gradshteyn} as
\begin{align}
&L\left[ {{f_{X_\varphi ^l}}\left( x \right)} \right]\left( s \right) = \sqrt \pi  \sum\limits_{u = 0}^\infty  {\sum\limits_{v = 0}^\infty  {\frac{{{{\left( {1 + \kappa } \right)}^{2\left( {u + 1} \right)}}{4^{u - v + 1}}}}{{{\kappa ^{ - u - v}}{{\left( {u!} \right)}^2}{{\left( {v!} \right)}^2}{e^{2\kappa }}}}} } {\rm{ }} \frac{{\Gamma \left( {2 + 2u} \right)\Gamma \left( {2 + 2v} \right)}}{{{{\left[ {s + 2\left( {\kappa  + 1} \right)} \right]}^{2u + 2}}\Gamma \left( {u + v + \frac{5}{2}} \right)}}\nonumber\\
&{ \times _2}{F_1}\left( {2 + 2u,\frac{1}{2} + u - v;\frac{5}{2} + u + v;\frac{{ - 2\left( {\kappa  + 1} \right) + s}}{{2\left( {\kappa  + 1} \right) + s}}} \right).
\end{align}

When $P_s^{act} \to \infty $, $s$ in the above equation goes to infinity. At the same time, the first term $\left( {i = 0,j = 0} \right)$ of above series dominates the whole expression, thus the Laplace transform is eventually simplified as
\begin{align}
{\cal L}\left[ {{f_{X_\varphi ^l}}\left( x \right)} \right]\left( s \right){ = _2}{F_1}\left( {2,\frac{1}{2};\frac{5}{2};1} \right)\frac{{16{{\left( {1 + \kappa } \right)}^2}}}{{3{e^{2\kappa }}{s^2}}}.
\end{align}

As $\sqrt {{X_\varphi }}  = \sum\nolimits_{l = 1}^L {\left| {X_\varphi ^l} \right|} $ and by applying the convolution theorem, the Laplace transform for the PDF of $\sqrt {{X_\varphi }}$  can be given by
\begin{align}
{\cal L}\left[ {f_{{\sqrt {{X_\varphi }} }}^{0 + }\left( x \right)} \right]\left( s \right)= {\left[ {_2{F_1}\left( {2,\frac{1}{2};\frac{5}{2};1} \right)\frac{{16{{\left( {1 + \kappa } \right)}^2}}}{{3{e^{2\kappa }}}}{s^{ - 2}}} \right]^L}.
\end{align}

After using the inverse Laplace transform, the above expression can be derived as
\begin{align}
f_{\sqrt {{X_\varphi }} }^{0 + }\left( x \right) = \frac{{{x^{2L - 1}}}}{{\left( {2L - 1} \right)!}}{\left[ {_2{F_1}\left( {2,\frac{1}{2};\frac{5}{2};1} \right)\frac{{16{{\left( {1 + \kappa } \right)}^2}}}{{3{e^{2\kappa }}}}} \right]^L}.
\end{align}

Furthermore, by applying equation ${F_\chi }\left( x \right) = \int_0^{\sqrt x } {{f_\chi }\left( x \right)dx} $, the approximated CDF of ${{X_\varphi }}$ at high SNRs can be finally expressed as
\begin{align}\label{H CDF}
F_{{X_\varphi }}^{0 + }\left( x \right) = \frac{{{\Lambda ^L}{x^L}}}{{{}\left( {2L} \right)!}}{\left[ {_2{F_1}\left( {2,\frac{1}{2};\frac{5}{2};1} \right)} \right]^L}.
\end{align}

Combining \eqref{H CDF} and \eqref{A3}, \eqref{The asymptotic pSIC} can be obtained after using Gauss-Laguerre quadrature and Gauss-Chebyshev quadrature.
The proof is complete.

\section*{Appendix~C: Proof of Theorem \ref{Theorem ER ipSIC}}\label{Appendix:C}
\renewcommand{\theequation}{C.\arabic{equation}}
\setcounter{equation}{0}
The PDF and CDF of $\gamma _r$ are first derived to facilitate the proof of the theorem. We note that the CDF of $\gamma _r$ can be obtained by transforming \eqref{ipSIC} as
\begin{small}
\begin{align}\label{C1}
{F_{{\gamma _r}}}\left( x \right) = \sum\limits_{k = 1}^K {\sum\limits_{u = 1}^U {\frac{{\pi \left( {{x_u}{\rm{ + }}1} \right){A_k}}}{{2U\Gamma \left( {{p_r}} \right)}}\sqrt {1 - x_u^2} } }  \gamma \left\{ {{p_r},\frac{{\sqrt x }}{{{q_r}}}\sqrt {\frac{{d_s^\alpha }}{{{a_r}P_s^{act}}}\left[ {\zeta \frac{{\sigma _s^2}}{{{\eta _0}}} + \frac{{\chi _u^\alpha }}{{\eta _0^2}}\left( {\frac{{\varepsilon {y_k}P_s^{act}}}{{{\beta _r}\lambda \sigma _{re}^{ - 2}}} + \frac{{\sigma _0^2}}{{{\beta _r}\lambda }}} \right)} \right]} } \right\},
\end{align}
\end{small}
By taking the derivative of above equation, we can obtain the PDF expression for $\gamma _r$ as
\begin{align}\label{C2}
{f_{{\gamma _r}}}\left( x \right) = \pi \sum\limits_{u = 1}^U {\sum\limits_{k = 1}^K {\frac{{\vartheta {A_k}\left( {{x_u}{\rm{ + }}1} \right)\sqrt {1 - x_u^2} }}{{2U\Gamma \left( {{p_r}} \right)2\sqrt x {e^{\vartheta \sqrt x }}}}} } {\left( {\vartheta \sqrt x } \right)^{{p_r} - 1}},
\end{align}
where $\vartheta  = \frac{1}{{{q_r}}}\sqrt {\frac{{d_s^\alpha }}{{{a_r}P_s^{act}}}\left[ {\zeta \frac{{\sigma _s^2}}{{{\eta _0}}} + \frac{{\chi _u^\alpha }}{{\eta _0^2}}\left( {\frac{{\varepsilon {y_k}P_s^{act}}}{{{\beta _r}\lambda \sigma _{re}^{ - 2}}} + \frac{{\sigma _0^2}}{{{\beta _r}\lambda }}} \right)} \right]} $.

By substituting \eqref{SINR n} into \eqref{efinition of ergodic rate}, the ergodic data rate expression of $U_r$ with ipSIC scheme is calculated as
\begin{align}\label{F gamma n}
R_{r,ipSIC}^{erg} = \left[ {{{\log }_2}\left( {1 + \frac{{{a_r}\lambda {\beta _r}P_s^{act}{{\left| {{\bf{h}}_s^H{{\bf{\Phi }}_r}{{\bf{h}}_r}} \right|}^2}}}{{\lambda {\beta _r}{{\left| {{\bf{n}}_s^H{{\bf{\Phi }}_r}{{\bf{h}}_r}} \right|}^2} + \varepsilon {{\left| {{h_{re}}} \right|}^2}P_s^{act} + \sigma _0^2}}} \right)} \right] = \frac{1}{{\ln 2}}\int_0^\infty  {\ln \left( {1 + x} \right){f_{{\gamma _r}}}\left( x \right)} dx.
\end{align}

By substituting \eqref{C2} into \eqref{F gamma n}, the expression of $U_r$'s ergodic data rate with ipSIC for ASTARS-NOMA networks can be given by
\begin{align}
R_{r,ipSIC}^{erg} = \int_0^\infty  {\sum\limits_{k = 1}^K {\sum\limits_{u = 1}^U {\frac{{\pi {A_k}\left( {{x_u}{\rm{ + }}1} \right)\ln \left( {1 + {x^2}} \right)\sqrt {1 - x_u^2} }}{{2U\ln 2\Gamma \left( {{p_r}} \right){e^{\vartheta x}}{\vartheta ^{ - {p_r}}}{x^{1 - {p_r}}}}}} } } dx.
\end{align}
Also by using Gauss-Laguerre quadrature, \eqref{ER ipSIC} can be obtained. The proof is complete.

\section*{Appendix~D: Proof of Theorem \ref{Theorem ER m}}\label{Appendix:D}
\renewcommand{\theequation}{D.\arabic{equation}}
\setcounter{equation}{0}
The expression for ${F_{{\gamma _t}}}\left( x \right)$ can be obtained by transforming \eqref{U_m} as
\begin{small}
\begin{align}\label{D1}
{F_{{\gamma _t}}}\left( x \right) = \sum\limits_{u = 1}^U {\gamma \left[ {{p_t},\sqrt {\frac{{xd_s^\alpha q_t^{ - 2}}}{{P_s^{act}\left( {{a_t} - x{a_r}} \right)}}\left( {\frac{{y_u^\alpha \sigma _0^2}}{{\eta _0^2{\beta _t}\lambda }} + \zeta \frac{{\sigma _s^2}}{{{\eta _0}}}} \right)} } \right]} \frac{{\pi \left( {{x_u}{\rm{ + }}1} \right)}}{{2U\Gamma \left( {{p_t}} \right)}}\sqrt {1 - x_u^2} .
\end{align}
\end{small}
Different from \eqref{C1}, the $ x$ in \eqref{D1} needs to satisfy the inequality $x < \frac{{{a_t}}}{{{a_r}}}$. Thus the ergodic data rate expression of $U_t$ is calculated as
\begin{align}\label{D2}
R_t^{erg} = \frac{1}{{\ln 2}}\int_0^{\frac{{{a_t}}}{{{a_r}}}} {{f_{{\gamma _t}}}\left( x \right)\ln \left( {1 + x} \right)} dx= \frac{1}{{\ln 2}}\int_0^{\frac{{{a_t}}}{{{a_r}}}} {\frac{1}{{1 + x}}\left[ {1 - {F_{{\gamma _t}}}\left( x \right)} \right]} dx.
\end{align}

By substituting \eqref{D1} into \eqref{D2}, the expression of $U_t$'s ergodic data rate for ASTARS-NOMA networks is given by
\begin{small}
\begin{align}\label{D3}
R_t^{erg} = \frac{1}{{\ln 2}}\int_0^{\frac{{{a_t}}}{{{a_r}}}} {\frac{1}{{1 + x}}} \left\{ {1 - \sum\limits_{u = 1}^U {\frac{{\pi \left( {{x_u}{\rm{ + }}1} \right)}}{{2U\Gamma \left( {{p_t}} \right)}}\sqrt {1 - x_u^2} } } \right.\left. {\gamma \left[ {{p_t},\frac{1}{{{q_t}}}\sqrt {\frac{{xd_r^\alpha }}{{\left( {{a_t} - x{a_r}} \right)P_s^{act}}}\left( {\frac{{y_u^\alpha \sigma _0^2}}{{\eta _0^2{\beta _t}\lambda }} + \zeta \frac{{\sigma _s^2}}{{{\eta _0}}}} \right)} } \right]} \right\}dx.
\end{align}
\end{small}

The definite integral in \eqref{D3} can be calculated by applying Gauss-Chebyshev quadrature. After some simple mathematical calculations, we can figure out \eqref{ER m}.
The proof is complete.

\bibliographystyle{IEEEtran}
\bibliography{mybib}

\end{document}